%% file: cumulant_encoder.tex
\documentclass[11pt]{article}
\usepackage{algorithm2e}
\usepackage{authblk}
\usepackage{amsmath}
\usepackage{amssymb}
\usepackage{braket}
\usepackage{graphicx}
\usepackage{ragged2e}
\usepackage{subcaption}
\usepackage{tikz}
\usepackage{tikz-3dplot}
\usepackage{verbatim}
\usepackage{xcolor}
\usepackage{braket}

\usepackage{hyperref}
\usepackage[numbers,sort&compress]{natbib}

\usetikzlibrary{positioning, fit}
\usetikzlibrary{shapes.geometric}

\tikzset{
    input/.style={draw, ellipse, thick, node distance=1cm, minimum height=0.8cm},
    hidden/.style={draw, rectangle, thick, node distance=1cm, minimum height=0.8cm},
    latent/.style={draw, ellipse, thick, node distance=1cm, minimum size=1cm},
    output/.style={draw, ellipse, thick, node distance=1cm, minimum height=0.8cm},
    redbox/.style={draw=black, line width=0.6mm, inner sep=4pt},
    bluebox/.style={draw=black, line width=0.6mm, dashed, inner sep=6pt},
}

\let\vec\boldsymbol
\newcommand{\x}{\vec x}

\newcommand{\y}{\vec y}

\newcommand{\tmi}{\tilde{\mathcal I}}

\newcommand{\beq}{\begin{eqnarray}}
\newcommand{\eeq}{\end{eqnarray}}

\begin{document}

\title{NCoder - A Quantum Field Theory Approach to Encoding Data}

\author[1]{D.S. Berman\thanks{d.s.berman@qmul.ac.uk}}
\author[2]{M.S. Klinger\thanks{marck3@illinois.edu}}
\author[1]{A.G. Stapleton\thanks{a.g.stapleton@qmul.ac.uk (Corresponding author)}}

\affil[1]{Centre for Theoretical Physics, Queen Mary University of London, London, E1 4NS, UK}
\affil[2]{Department of Physics, University of Illinois, Urbana IL 61801, USA}

\maketitle

\begin{abstract}
In this paper we present a novel approach to interpretable AI inspired by Quantum Field Theory (QFT) which we call the \emph{NCoder}. The NCoder is a modified autoencoder neural network whose latent layer is prescribed to be a subset of $n$-point correlation functions. Regarding images as draws from a lattice field theory, this architecture mimics the task of perturbatively constructing the effective action of the theory order by order in an expansion using Feynman diagrams. Alternatively, the NCoder may be regarded as simulating the procedure of statistical inference whereby high dimensional data is first summarized in terms of several lower dimensional summary statistics (here the $n$-point correlation functions), and subsequent out-of-sample data is generated by inferring the data generating distribution from these statistics. In this way the NCoder suggests a fascinating correspondence between perturbative renormalizability and the sufficiency of models. We demonstrate the efficacy of the NCoder by applying it to the generation of MNIST images, and find that generated images can be correctly classified using only information from the first three $n$-point functions of the image distribution. 
\end{abstract}

\tableofcontents

\newpage

\section{Introduction}%
\label{sec:bayesian_renormalization}

Neural networks have become ubiquitous in modern data science, and for good reason. As universal function approximators, neural networks can produce effective outputs in a wide array of different learning and optimization tasks. Unfortunately, the universal applicability of neural networks comes at a cost -- exceptionally large numbers of parameters, and implicit non-linearities. These and other characteristics of neural networks render them very much like the prototypical black box -- great for solving problems but practically useless for understanding the reasoning behind the solution. 

In previous work \cite{Berman:2022mak,Berman:2022uov,Berman:2023rqb} the authors of this note have initiated a course of study aimed at improving the interpretability of neural networks, largely by appealing to ideas from an area of theoretical physics known as the renormalization group (RG). The basic premise of RG is that the degrees of freedom constituting a quantum or statistical field theory can be organized according to a hierarchy of physical scales such that some degrees of freedom impact physics only at energy scales beyond a particular cutoff. In this way, a physicist who is only capable or interested in observing a system at a particular energy scale may coarse grain the complete description by systematically removing all those modes which are ``irrelevant" to outcomes at the observed scale. 

In its conventional form, RG relies rather heavily on assumptions about the locality of interactions in the system being studied in order to construct the requisite hierarchy of scales mentioned above. For this reason, it can be a challenge to see how the reasoning of renormalization may be outsourced into data scientific tasks where such assumptions are almost certainly not met. With that being said, in \cite{Berman:2023rqb} it was demonstrated that, although a statistical model typically will not possess any native notion of locality, the family of statistical models which could possibly describe a particular system comes equipped with a natural emergent notion of locality through the concept of information geometry \cite{amari2016information,nielsen2020elementary}. Leveraging this observation, \cite{Berman:2023rqb} illustrated how one can ``renormalize" a simple autoencoder neural network, and demonstrated that a majority of the parameters which define the network could be removed without qualitatively or quantitatively affecting its performance.

In this note we build on previous work by again appealing to intuitions from physics to guide our approach towards more interpretable neural networks.\footnote{There have been many recent works exploring the intersections between Machine Learning techniques and high-energy theoretical physics \cite{decelle2023inferring, Dubey:2023dvu,Cole:2021nnt, Erbin:2022lls, Halverson:2023ndu, Yip:2023vud, Anderson:2023viv, Berglund:2023ztk, nielsen2020elementary, Niarchos:2023lot, Kantor:2022epi, Cotler_2023, cotler2023renormalizing, bachtis2023inverse, aarts2021interpreting, deluca2023improving}, with some even exploring an ML approach to correlation functions, including \cite{PhysRevResearch.4.043143, decelle2023inferring}.}  In quantum field theory a natural set of observables are obtained in terms of $n$-fold products of the fields. The expectation values of such observables define what are called $n$-point correlation functions, or simply $n$-point functions. The most common computation in a field theory is of matrix elements which encode the transition probability for events involving the interaction of various fields. These matrix elements can be formulated as a perturbation series written in terms of the aforementioned $n$-point functions. The resulting matrix elements are organized so that, roughly speaking, higher $n$-point functions appear at higher orders in the perturbation series. 

In parlance more familiar to data science, $n$-point correlation functions are the moments of the quantum field theory when regarded as a probability model in the space of field configurations. These moments can subsequently be reorganized into cumulants, which are referred to as connected or irreducible correlation functions in the physics community. From this perspective, the role of the $n$-point functions in framing a perturbative analysis as described above is not surprising. The data encoded in a general probability distribution can equally well be encoded in either a moment generating or cumulant generating functional which has the appearance of a Taylor expansion in the $n$-point functions of the theory. In this regard, incorporating knowledge about higher $n$-point functions resolves finer detail in the probability model in the same way that adding more terms to a Taylor expansion allows for a reconstructed function to account for more fine grained features encoded in higher derivatives. 

More formally, the relationship between a probability distribution and its moments\footnote{As discussed in the main text, the reconstruction problem can also be formulated with respect to cumulants. For simplicity we will simply refer to the moments in this paragraph.} can be understood in terms of the moment reconstruction problem. As it is typically formulated, one is given the set of all moments of a particular probability distribution and subsequently seeks to deduce the distribution itself. As a modification of the moment reconstruction problem, one can seek to build a probability measure by matching to some subset of all its moments. In this modified problem, the solution to the moment reconstruction problem will certainly not be unique. Rather, it will refer to an equivalence class of probability measures whose moments agree on the prescribed subset, but which can differ relative to any moments which have not been specified. In this respect we once again see a sense in which the specification of higher moments improves the resolution of the probability measure by breaking the degeneracy caused by incompletely specifying the moment reconstruction problem. This observation can be formalized in terms of the Edgeworth expansion, which is introduced in Section \ref{sub:_the_hamburger_problem}.

The goal of this note is to investigate the role of $n$-point functions in organizing the structure of neural networks. Put differently, we are interested in developing and testing the validity of perturbative analyses for neural networks inspired by quantum field theory. In particular, we ask the question -- how does the performance of a neural network change with the inclusion of higher $n$-point data into its training sample? In other words, we are engaging in a sequence of incomplete moment reconstruction problems as described above. If a neural network shares the hierarchical structure of a quantum field theory, it should be the case that the network accesses finer and finer features of the sample as the higher $n$-point functions are included. In fact, this structure is very evocative of the neural network-field theory correspondence explicated in \cite{Halverson:2020trp,Maiti:2021fpy,Halverson:2021aot,Demirtas:2023fir}. 

A natural extension of our question is whether or not the `perturbative' reconstruction of the probability distribution can be truncated at finite order. That is, can the data generating distribution be resolved (up to a chosen level of accuracy) in terms of a finite, computationally feasible number of $n$-point correlation functions? We note that this question bears a resemblance to the question of perturbative renormalizability in QFT. A field theory is termed perturbatively renormalizable if there exist a \emph{finite} number of parameters which can be tuned so that the ultraviolet divergences of the theory are canceled; this will be discussed in more detail in the discussion section.


To address our various questions we introduce in this note the \emph{NCoder} -- a neural network built around our aim for interpretability. The NCoder can be viewed as a deconstructed autoencoder in which the `latent layer' is not built during the simultaneous optimization of the encoder/decoder but rather is fixed before training. Although this constraint limits the freedom of the network, it improves its `readability'. Motivated by the central role played by $n$-point functions in the analysis of field theories and general data science models, the NCoder uses the space of correlation functions as its prescribed latent layer. This set up renders the NCoder a potentially powerful tool for analyzing the importance of interpretable statistics in data generation and classification tasks. The paper is organized as follows: in Section \ref{sec:statistical_prerequisites_and_cumulant_representations} we review relevant concepts from probabiltiy and statistics such as the definition of moments and cumulants. In Section \ref{sub:_the_hamburger_problem} we introduce the moment/cumulant reconstruction problem, and present a general solution in terms of the Edgeworth expansion which generalizes ideas from perturbation theory to arbitrary probability models. In Section \ref{sec:images_as_lattice_realizations} we briefly review how image data can be regarded as draws from a lattice field theory in preparation for the analysis of the latter sections. Section 
\ref{sec:reconstruction} presents the NCoder in detail, with a special focus on rendering the task of obtaining $n$-point functions feasible in light of the computational cost of the combinatorics involved. In Section \ref{sec:experimental_setup} we use the NCoder to perform image generation tasks involving the MNIST \cite{deng2012mnist} and Fashion MNIST \cite{DBLP:journals/corr/abs-1708-07747} datasets. We conclude in Section \ref{sec:discussion} in which we discuss our findings, and suggest interesting directions for future work.

\section{Statistical prerequisites}%
\label{sec:statistical_prerequisites_and_cumulant_representations}

In this section we review some pertinent ideas from probability theory and statistics. In particular, we introduce moments and cumulants for a probability distribution, and discuss a perturbative approach to reconstructing the distribution from this data. A useful reference on this subject is \cite{mccullagh2018tensor}, a more standard reference is \cite{wasserman2004all}.

\subsection{Moments, cumulants, and generating functions}

Let $(X,\mu)$ be a measure space and denote by $L^1(X,\mu)$ the set of $\mu$-measurable functions on $X$.\footnote{The measure $\mu$ is sometimes referred to as a reference measure, and defines a notion of integration on $X$. In most cases it is sufficient to regard $\mu$ as the standard Lebesgue measure \cite{reed2012methods}.}  We regard $X$ as encoding the data of a random variable whose statistics are determined by placing some distribution over the space. A probability distribution on $X$ is a measurable function $p \in L^1(X,\mu)$ which is normalized in the sense that
\beq
	1 = \int_{X} \mu(x) p(x). 
\eeq
Given a function $f \in L^1(X,\mu)$ we introduce the notation
\beq
	\mathbb{E}_p(f) \equiv \int_{X} \mu(x) p(x) \; f(x), 
\eeq
which encodes the expectation value of $f$ with respect to the distribution $p$.

Suppose that $X$ has the structure of an $n$-dimensional vector space so that $X \ni x = (x^1, ..., x^n)$. Then, we define the $n^{th}$ moment of the distribution $p$ as given by
\beq \label{Moments}
	\mu_{p}^{i_1 ... i_n} \equiv \mathbb{E}_p(x^{i_1} \cdot \cdot \cdot x^{i_n}) = \int_{X} \mu(x) p(x) \; x^{i_1} \cdot\cdot\cdot x^{i_n}. 
\eeq 
The moment generating function associated with $p$ is
\beq \label{Moment Generating Function}
	M_{p}(\xi) \equiv \mathbb{E}_p(e^{\xi(x)}),
\eeq
where $X^* \ni \xi = (\xi_1, ..., \xi_n)$ is an element of the dual space and the pairing $\xi(x) = \xi_i x^i$. The moment generating function is so named because if one expands the exponential appearing in \eqref{Moment Generating Function} one finds\footnote{Here, we've used the linearity of the expectation.}
\beq \label{Moments 2}
	M_{p}(\xi) = \mathbb{E}_p\bigg(\sum_{n = 0}^{\infty} \frac{1}{n!} \xi_{i_1} \cdot\cdot\cdot \xi_{i_n} x^{i_1} \cdot\cdot\cdot x^{i_n} \bigg) = \sum_{n = 0}^{\infty} \frac{1}{n!} \xi_{i_1} \cdot\cdot\cdot \xi_{i_n} \mu_{p}^{i_1 ... i_n}. 
\eeq
One may therefore compute the moments of $p$ by taking derivatives of the moment generating function:
\beq
	\mu_{p}^{i_1 ... i_n} = \frac{\partial^n M_{p}}{\partial \xi_{i_1} \cdot\cdot\cdot \partial \xi_{i_n}}\bigg\rvert_{\xi = 0}. 
\eeq

Related to the moment generating function is the cumulant generating function:
\beq \label{Cumulant Generating Function}
	K_{p}(\xi) \equiv \ln M_{p}(\xi).
\eeq
The cumulants of the distribution $p$, $\kappa_{p}^{i_1 ... i_n}$, are \emph{defined} in terms of the cumulant generating function such that
\beq \label{Cumulants}
	K_{p}(\xi) \equiv \sum_{n = 1}^{\infty} \frac{1}{n!} \xi_{i_1} \cdot\cdot\cdot \xi_{i_n} \kappa_{p}^{i_1 ... i_n},
\eeq
or in other words
\beq
	\kappa^{i_1 ... i_n}_{p} = \frac{\partial^n K_{p}}{\partial \xi_{i_1} \cdot\cdot\cdot \partial \xi_{i_n}}\bigg\rvert_{\xi = 0}. 
\eeq
Inserting \eqref{Moments 2} and \eqref{Cumulants} back into \eqref{Cumulant Generating Function} and equating both sides term by term as polynomials in $\xi$ we can obtain a relationship between the moments of the distribution $p$ and its cumulants. 

In particular, let $\mathcal{P}_n$ denote the set of partitions of $n$ elements. Recall that a partition $\sigma \in \mathcal{P}_n$ consists of a collection of non-empty subsets of $[n] \equiv \{1, ..., n\}$ called blocks such that each pair of blocks is disjoint and the union over all of the blocks is equal to $[n]$. Given a multi-index $I = (i_1 ... i_n)$ and a subset $b \subset [n]$ let us introduce the notation:
\beq
	\mu^{I_b}_{p} \equiv \mathbb{E}_p\bigg(\prod_{k \in b} x^{i_k}\bigg).
\eeq	
Similarly, let $\kappa^{I_b}_{p}$ denote the cumulant associated with the set $\{x^{i_k}\}_{k \in b}$ regarded as a family of jointly distributed random variables with distribution $p$. Then, we have the following pair of equations which encode the relation between moments and cumulants\footnote{In what follows $I = (i_1 ... i_n)$ is an n-tuple multi-index.}
\beq \label{Moment-Cumulant Formulae}
	\mu_{p}^{I} = \sum_{\sigma \in \mathcal{P}_n} \prod_{b \in \sigma} \kappa^{I_b}_{p}, \qquad \kappa^{I}_{p} = \sum_{\sigma \in \mathcal{P}_n} (-1)^{\# \sigma - 1}(\# \sigma-1)! \prod_{b \in \sigma} \mu_{p}^{I_b}. 
\eeq
In \eqref{Moment-Cumulant Formulae} we have used the standard notation $\#\sigma$ to refer to the number of blocks in the partition $\sigma$.\footnote{As a sanity check, consider the problem of computing the second cumulant in terms of the moments. There are two partitions of $[2]$:  $12$ and $1 | 2$. Plugging these back into the formula \eqref{Moment-Cumulant Formulae} we find
\beq
	\kappa_{p}^{i_1 i_2} = \mu_{p}^{i_1 i_2} - \mu_{p}^{i_1}\mu_{p}^{i_2}. 
\eeq} 

\subsection{The reconstruction problem and the Edgeworth expansion}%
\label{sub:_the_hamburger_problem}

A classic problem in the field of statistics is the `moment problem' or `Hausdorf moment problem' \cite{schmudgen2017moment,lin2017recent,schmudgen2020ten}. Roughly speaking, the moment problem asks whether one can reconstruct the probability distribution $p$ using the data provided by its moments, $\mu_p^{I}$, or some subset therein. That is, given a set $\{\mu^I\}_{I \in \mathcal{I}}$, where $\mathcal{I}$ is some collection of multi-indices, we seek a probability distribution $p$ such that
\begin{equation}
	\int_{X} \mu(x) p(x) \; \prod_{i \in I} x^i = \mu^I, \; \forall I \in \mathcal{I}. 
\end{equation}
Although it is less frequently studied, one may also consider a version of the reconstruction problem in which cumulants are presented instead of moments. This is referred to as the cumulant reconstruction problem. 

In the following sections, we present an approximate solution to the moment/cumulant problem in which a neural network is trained to reconstruct the distribution of an underlying data set using a truncated set of correlation functions. Before moving on to the numerics, however, let us introduce an analytic approach to the reconstruction problem which is well known in the mathematics literature. 

We again take $(X,\mu)$ to be a measure space. Suppose that $p$ and $p_0$ are a pair of distributions in $L^1(X,\mu)$. We will now review a general formalism for obtaining a distribution $p$ as a perturbation series starting from the distribution $p_{0}$ along with the input of the cumulants of $p$. This approach is known as the Edgeworth expansion \cite{hall2013bootstrap,hall1983inverting,Demirtas:2023fir}. Let us define
\beq \label{Relative cumulants}
	\eta^{I} \equiv \kappa_{p}^{I} - \kappa_{p_0}^I 
\eeq
which is the difference between the cumulants of $p$ and $p_0$. The difference in the cumulant generating functions is therefore given by
\beq \label{Relative cumulant generating}
	K_{p}(\xi) - K_{p_0}(\xi) = \sum_{n = 1}^{\infty} \frac{1}{n!} \xi_{i_1} \cdot\cdot\cdot \xi^{i_n} \eta^{i_1 ... i_n}. 
\eeq	
Using \eqref{Cumulant Generating Function}, and exponentiating both sides of \eqref{Relative cumulant generating} we obtain the equation
\beq \label{Relative moment generating}
	\frac{M_{p}(\xi)}{M_{p_0}(\xi)} = \text{exp}\bigg(\sum_{n = 1}^{\infty} \frac{1}{n!} \xi_{i_1} \cdot\cdot\cdot \xi^{i_n} \eta^{i_1 ... i_n}\bigg) \equiv \sum_{n = 0}^{\infty} \frac{1}{n!} \xi_{i_1} \cdot\cdot\cdot \xi_{i_n} \delta^{i_1 ... i_n}. 
\eeq
In the final equality of \eqref{Relative moment generating} we've used \eqref{Moment-Cumulant Formulae} to define the ``moments" $\delta^{I}$ associated with the ``cumulants" $\eta^I$.\footnote{We should note, however, that $\delta^I$ needn't be the moments of any analytic distribution. Moreover $\delta^I \neq \mu_{p}^I - \mu_{p_0}^I$.}

Rewriting \eqref{Relative moment generating} as 
\beq \label{Relative moment 2}
	M_{p}(\xi) = M_{p_0} \sum_{n = 0}^{\infty} \frac{1}{n!} \xi_{i_1} \cdot\cdot\cdot \xi_{i_n} \delta^{i_1 ... i_n},
\eeq	
we can see that it provides an expression for the moment generating function of $p$ in terms of that of $p_0$. We can ``invert" \eqref{Relative moment 2} term by term by recognizing that
\beq \label{Integration by parts}
	\xi_{i_1} \cdot\cdot\cdot \xi_{i_n} M_{p_0}(\xi) = (-1)^n \int_{X} \mu(x) \frac{\partial^n p_{0}}{\partial x^{i_1} \cdot\cdot\cdot \partial x^{i_n}} e^{\xi(x)}. 
\eeq
Here, we have used integration by parts and the fact that $p_{0}$ is a normalized and hence compactly supported function. Plugging \eqref{Integration by parts} into \eqref{Relative moment 2} and expanding the definition of the moment generating function on the left hand side we obtain the equation
\beq \label{Relative moment 3}
	\int_{X} \mu(x) p(x) e^{\xi(x)} = \int_{X} \mu(x) \bigg(\sum_{n = 0}^{\infty} \frac{(-1)^n}{n!} \delta^{i_1 ... i_n} \frac{\partial^n p_{0}}{\partial x^{i_1} \cdot\cdot\cdot \partial x^{i_n}} \bigg) e^{\xi(x)}. 
\eeq
Thus, we conclude that
\beq \label{Edgeworth Expansion}
	p(x) = \sum_{n = 0}^{\infty} \frac{(-1)^n}{n!} \delta^{i_1 ... i_n} \frac{\partial^n p_{0}}{\partial x^{i_1} \cdot\cdot\cdot \partial x^{i_n}}\bigg\rvert_{x}. 
\eeq

In what follows we will often work under the assumption that $p_0$ is a Gaussian distribution whose first and second cumulants match with those of $p$. In this case \eqref{Edgeworth Expansion} takes the form
\beq \label{Edgeworth Normal}
	p(x) = p_0(x) + \sum_{n = 3}^{\infty} \frac{(-1)^n}{n!} \mu_p^{i_1 .. i_n} \frac{\partial^n p_0}{\partial x^{i_1} \cdot\cdot\cdot \partial x^{i_n}}\bigg\rvert_{x},
\eeq
where now $\mu_p^{I}$ are genuinely the moments of the distribution $p$. In words \eqref{Edgeworth Normal} states that the distribution $p$ can be approximated by a Gaussian plus corrections whose significance is governed by the magnitude of the $n$-moments of $p$. As was noted in \cite{Demirtas:2023fir}, this resembles closely the structure of a perturbative expansion from quantum field theory \cite{peskin2018introduction}. Our goal is to test whether \eqref{Edgeworth Normal} encodes a sense in which Neural Networks learn hierarchically by extracting finer information from higher moments in the duration of training.



\section{Images as lattice realizations}
\label{sec:images_as_lattice_realizations}

In this note we will be interested in applying machine learning techniques to image generation tasks. In light of the previous section, we would like to develop a probability theoretic understanding of this problem. The approach we will take is to regard a collection of images as realizations from a lattice field theory. That is, we treat each image as a random variable coinciding with the set of pixel realizations at each point on a discrete lattice. 

To be more concrete, let $L$ be a lattice of points. For our purposes we take $L$ to constitute pixel sites $x \in L$ in an image of dimension $L_x \times L_y$; with $L_x$ the number of horizontal pixels, and $L_y$ the number of vertical pixels. For definiteness, we will index the lattice points as $x_i$ such that $i$ is ordered with the bottom left lattice site associated with the $0$th index, and $i$ increases left-to-right, bottom-to-top (as in Figure \ref{fig:map_lattice}) -- this concrete distinction allows element-wise operations to be defined more absolutely in following sections. We also introduce the notation $|L| = L_x L_y$ to denote the total size of the lattice. 

A lattice field is a map 
\begin{equation}
	\label{eqn:lattice_sample_space_map} \Phi: L \to S
\end{equation}
\begin{figure}[tpb]
	\begin{center}
		\makebox[\linewidth]{\input{figs/tikz/map_lattice}}
		\caption{Mapping between the lattice sites $x_i \in L$ and the associated
		random variables in the sample space $y_i = \Phi(x_i) \in S$.}%
		\label{fig:map_lattice}
	\end{center}
\end{figure}
which assigns, point-wise, an element $y_i := \Phi(x_i)$ in the sample space $S$. To simplify notation, we define $\Phi_i := \Phi(x_i)$ to be the realized output at the $i$-th lattice site in the sample space in which case we may regard the full lattice field simply as a large vector\footnote{Bear in mind that each $\Phi_i$ is itself an element of $S$ and therefore may be vector valued.}
\begin{equation} \label{Lattice Field}
	\Phi(x) \equiv \big(\Phi_0, \ldots,  \Phi_{|L|-1} \big).
\end{equation}

In what follows we will often take $S = [L_c]$ in which case \eqref{Lattice Field} describes the pixels of an image that may take $L_c$ discrete pixel values. As we have alluded to, we will treat $\Phi$ as a random variable drawn from some distribution. That is $\Phi$ is described by $p(\Phi)$ which should be interpreted as a joint distribution over all of the components in \eqref{Lattice Field}. In the case that $S$ is a discrete sample space
\begin{equation} 
	\label{eqn:prob_dist_pixels}
	p(y) = \text{Pr}(\Phi_0 = y_0, ..., \Phi_{|L|-1} = y_{|L|-1})
\end{equation}
is the probability that $\Phi(x_i)$ takes the value $y_i \in S$ across all lattice sites $i$.

Given the distribution $p$ governing lattice realizations, we can compute moments and cumulants in the same way as was described in Section \ref{sec:statistical_prerequisites_and_cumulant_representations}. For example,
\beq
\label{eqn:moment_of_full_dataset} 
	\mu_p^{i_1 ... i_n} \equiv \mathbb{E}_p\bigg(\Phi(x_{i_1}) \cdot\cdot\cdot \Phi(x_{i_n})\bigg).
\eeq
Notice that this $n$-point function may be interpreted as a function on $L^{\times n}$, for this reason we may also denote it by
\beq
	\mu_p(x_{i_1}, ..., x_{i_n}) \equiv \mu_p^{i_1 ... i_n},
\eeq 
to stress this point. The cumulants $\kappa^{i_1, \ldots, i_n}_p \equiv \kappa_p(x_{i_1}, ..., x_{i_n})$ are subsequently obtained via the moment-cumulant formulae \eqref{Moment-Cumulant Formulae}. In this context the moments $\mu_p(x_{i_1}, ..., x_{i_n})$ and cumulants $\kappa_p(x_{i_1}, ..., x_{i_n})$ are the lattice analogs of, respectively, disconnected and connected correlation functions which play a central role in the study of continuum quantum and statistical field theories, see Appendix \ref{sec:_cumulants_in_quantum_field_theory}. 

For future reference, we introduce the following notation to refer to collections of moments and cumulants:
\beq \label{Moment and Cumulant sets}
	\mathcal{M}^{\mathcal{I}} \equiv \{\mu^{I} \; | \; I \in \mathcal{I} \}, \qquad \mathcal{K}^{\mathcal{I}} \equiv \{\kappa^{I} \; | \; I \in \mathcal{I}\}. 
\eeq
Here, $\mathcal{I}$ is and index set whose elements are multi-indices $\mathcal{I} \ni I = (i_1 ... i_n)$.\footnote{The multi-indices in $\mathcal{I}$ needn't be of the same length.} In certain special situations a truncation of the notation introduced in \eqref{Moment and Cumulant sets} is warranted. In particular, we let
\beq \label{Moment and Cumulant sets 2}
	\mathcal{M}^{(k_i,k_f)}, \qquad \mathcal{K}^{(k_i,k_f)}
\eeq 
denote the sets of all moments/cumulants of order $k_i$ through order $k_f$.\footnote{The order of a cumulant is equal to the order of the highest moment that appears in its definition via \eqref{Moment-Cumulant Formulae}} It is sufficient to consider only the \textit{combinations with replacement} at each order $n$ as opposed to \textit{permutations} due to the $S_n$ symmetry in the moment-cumulant expansion.\footnote{By combinations with replacement we mean draws of items from a set of identifiable members, such that the selection order does not matter (c.f. permutations); items are replaced after being drawn.}

In general it is intractable to compute an analytic expression for the distribution $p$ governing lattice realizations, especially when the number of lattice sites becomes large. However, one may construct
a sampling or approximation method (for example a neural network) to create an effective description. In the following sections we propose a representation
of this data inspired by quantum field theory based around learning the distribution $p$ by observing its $n$-point functions. Qualitatively, what we will do is to encourage a neural network to compute the distribution $p$ using the Edgeworth expansion by sequentially incorporating higher $n$-point functions into the training.

\section{Encoding and decoding samples from $n$-point functions}%
\label{sec:reconstruction}
We now turn to the main analysis of the paper. The problem we are interested in addressing is how one can construct interpretable encodings of image data. To accomplish this goal we will take inspiration from the moment problem and the Edgeworth expansion \eqref{Edgeworth Expansion}. The basic idea is to use collections of $n$-point correlation functions as the latent layer for an image autoencoder. We note that strictly speaking the aforementioned construction is not an autoencoder due to the fixed latent space, however henceforth we will use the term `n-point autoencoder', or NCoder for short, to mean `n-point correlation space encoder-decoder'. Regarding images as realizations of a lattice field theory, higher $n$-point functions qualitatively encode departures from free field theory. More generally, these higher point functions encode the departure of the data generating distribution for the images from that of a Gaussian random process (GRP). In this sense an autoencoder which utilizes data from higher $n$-point functions may be interpreted as performing a perturbative analysis of some complicated interacting field theory.

To pinpoint the role of higher $n$-point functions in resolving finer image detail we organize our analysis order by order in the $n$-point functions admitted into the latent layer of the autoencoder. That is, we begin by training a model in which the latent layer consists only of $1$-point realizations from a sample of image data, then proceed to a model with both $1$- and $2$- point functions and so on. For reasons of computational cost we terminate this analysis at $3$-point data. In the language of field theory this implies that we are considering leading order corrections above the best free-field approximation to the theory underlying image realizations.  


\subsection{Review of autoencoders}%
\label{ssub:review_of_autoencoders}
An autoencoder is a neural network which is often used for the dimensional reduction of samples from an input dataset \cite{baldi2012autoencoders,schmidhuber2015deep,doersch2016tutorial} (these citations are clearly not exhaustive). Autoencoders learn efficient representations of message space samples $y$ by encoding through a lower-dimensional bottleneck and decoding back to the original message space. Data is passed through an encoder of hidden layers which transform it into a lower dimensional representation whilst aiming to retain the most salient features of the source dataset. As such, the loss function is typically chosen such that it is extremized when the most important features of the data are retained. From the latent space representation, a decoder attempts to reconstruct the source data whilst minimizing the autoencoder loss between the input and the output of the network. 

More formally, let the encoding and decoding spaces be Euclidean, such that
\begin{align}
	\chi &\simeq  \mathbb R^n & \eta &\simeq \mathbb R^m,
\end{align}
where $n$ is the dimensionality of the input data, and $m$ the dimension of the \textit{encoded space} or \textit{latent space} chosen when the network is initialized. For each data instance $y \in \chi$, there is a corresponding representation in the latent space which we denote by $\ell \in \eta$. 

The architecture of the autoencoder corresponds to the specification of a pair of mappings. First, we have the \emph{encoder function} $E_\theta$, parameterized by a set of variables $\{\theta_i \}$, which maps from the source message space $\chi$ to the latent space $\eta$, 
\begin{equation} \label{Encoder Function}
	E_\theta: \chi \to \eta.
\end{equation}
Similarly, we have the \textit{decoder function} $D_\phi$, parameterized by variables $\{ \phi_i \}$, which maps back from the latent space into the source message space:
\begin{equation}
	D_\phi: \eta \to \chi.
\end{equation}

Conventionally, autoencoders are used to find the most compact representation of data samples $y$ through an information bottleneck \cite{tishby2000information} (here the bottleneck is the latent space). The network is thus trained by minimising the \textit{autoencoder loss},
\begin{equation}
    \label{eqn:single_autoencoder_loss}
    \mathcal L(y | \theta,\phi) = \frac{1}{n} \sum_{j=1}^n  f(|y_j - D_\phi E_\theta(y)_j|),
\end{equation}
where $f(x)$ is some monotonically decreasing function which is minimized at $x = 0$.

As an explicit example, one can think of a toy feed-forward autoencoder with 2 fully connected linear layers and a ReLU activation function between the layers. The first layer encodes samples into a latent space, and the second acts as a decoder mapping back to the original sample space\footnote{This is a similar architecture to a Restricted Boltzmann Machine, albeit with a loss function which is not a free-energy \cite{hinton2012practical}.}. 
A diagram of the model is shown in Figure \ref{fig:simple_autoencoder}.

\begin{figure}[ht]
	\centering
	\input{figs/tikz/reconstruction.tex}
	\caption{Simple fully connected encoder-decoder architecture for the case
	of 4d input data and 4d output data.}
	\label{fig:simple_autoencoder}
\end{figure}

\paragraph{Collection autoencoders}%
\label{par:collection_autoencoders}

Having established the basic features of an autoencoder, we will now aim to introduce a slight generalization which will be necessary for our implementation. Instead of taking an encoder which maps from a single copy of the message space, it will be necessary from our perspective to consider a case where the encoder maps from products of multiple copies of the message space. To this end, we again let $E_\theta$ denote an encoder function parameterized by variables $\{\theta_i\}$, however, unlike in \eqref{Encoder Function}, $E_\theta$ now maps from $m_\text{in}$ copies of the message space into the latent space $\eta$,
\begin{equation}
	\label{eqn:collection_encoder} 
	E_\theta: \underbrace{\chi \times \ldots \times \chi}_{m_\text{in} \text{ times}} \to \eta.
\end{equation}
Similarly, we may augment the corresponding decoder $D_{\phi}$, parameterized by variables $\{\phi_i\}$, to map back into $m_\text{out}$ copies of the message space:
\begin{equation}
	\label{eqn:batch_decoder} 
	D_\phi: \eta \to \underbrace{ \chi \times \ldots \times \chi}_{m_\text{out} \text{ times}}.
\end{equation}
In most cases $m_\text{in} \geq m_\text{out}$. For our purposes, an interesting use of a collection autoencoder is to perform a `purification' of the input samples in order to produce platonic\footnote{By `platonic' we mean `sharing the salient features with the majority of samples' or \textit{quintessential}.} outputs. To be more explicit, a collection autoencoder will extract salient features from the batch of input samples and produce a sample which is representative of the input distribution. For a sufficiently well-trained model, one may consider this as a type of averaging procedure, where the averaging occurs over the features of the distribution that the network learns during training. This is done by taking the size of the input space to be strictly larger than that of the output space, $m_\text{in} > m_\text{out}$, so that the autoencoder is forced to perform some averaging over the features of the input sample. One may implement this most simply in practice by modifying the loss of \eqref{eqn:single_autoencoder_loss} such that $\mathcal L(y|\theta, \phi) \to \mathcal N \sum_\beta \mathcal L^\beta(y|\theta, \phi)$, where $\beta$ indexes each element in the collection. In the ensuing sections we will frequently take advantage of this idea. 

\subsection{NCoder -- $n$-point autoencoders}%
\label{sub:ncoder}

One common property of the autoencoders presented in Section \ref{ssub:review_of_autoencoders} is that the encoder and decoder are trained simulataneously; that is the network's optimiser updates the parameters of both the encoder and decoder functions in each iteration. A benefit of this approach is that the latent space is determined completely by the network, and is thus not free to be determined a-priori. At the same time, this can be disadvantageous in the sense that the user has little to no intuition as to how the autoencoder chooses to encode samples -- autoencoders are \textit{`black-box'} systems.

In this note we introduce a novel autencoder construction with the intention of getting under the hood and understanding more about how complicated data can be efficiently encoded. We are interested in the case where the message space consists of images regarded as outputs from a lattice field theory as discussed in Section \ref{sec:images_as_lattice_realizations}. Rather than leaving the construction of the latent space up to the model optimizer, we \emph{require} the encoder to map collections of samples from the message space to a restricted set of sampled $n$-point functions, whilst the decoder analogously maps from the space of $n$-point functions back to (products of) the message space. To enforce the fixing of the latent space the $n$-point autoencoder will train the encoder and decoder functions separately. 
 
For definiteness, consider an adaptation of \eqref{eqn:collection_encoder} and \eqref{eqn:batch_decoder} where the latent space is fixed to be the space\footnote{Here we have chosen to encode $n$-point correlation data via cumulants, but one could have alternatively selected moments. Cumulants have the nice feature that $n$-point and $m$-point cumulants encode distinct correlation information for $n \neq m$. In the physics context cumulants are said to encode `irreducible correlations' for this reason.} $\mathcal{K}^{\tilde{\mathcal{I}}}$ as defined in \eqref{Moment and Cumulant sets},
\begin{align}
	\label{eqn:cumulant_autoencoder_spaces_unmasked} 
	E_{\theta}: \chi^{m_\text{in}} &\to \mathcal K^{\tilde{\mathcal{I}}}  & D_{\phi}: \mathcal K^{\tilde{\mathcal{I}}} &\to \chi^{m_\text{out}},\\
	y_\text{in} &\mapsto K^{\tilde{\mathcal{I}}} & K^{\tilde{\mathcal{I}}} &\mapsto z_\text{out},
\end{align}
where $\tilde{\mathcal I}$ is a subset of the full index set associated with all $n$-pt functions included in \eqref{Moment and Cumulant sets}.
Here $(y_\text{in})_r$ is a vector of flattened\footnote{By `flattened' we mean a vector which combines the sample index $j$ and the lattice site index $i$; this is merely introduced for computational convenience.} input samples, and $K^{\tilde{\mathcal{I}}}$ is a vector of the associated $n$-point functions with respect to collections of lattice points contained in the index set $\tilde{\mathcal{I}}$. The flattening process simply vectorizes the training set such that
\begin{equation}
	\label{eqn:flattened_training_data} 
	y_\text{in} = (T^{0}_0, \ldots, T^{j}_{i}, \ldots, T^{|T|}_{|L|}),
\end{equation}
where $i$ indexes the lattice sites and $j$ indexes the sample number in the set. 

We would like to stress the following interpretation for the autoencoder defined by \eqref{eqn:cumulant_autoencoder_spaces_unmasked}. Firstly, the encoder is learning how to encode data in terms of their associated $n$-point correlation functions. Secondly, the decoder is learning how to reproduce data given the information contained in this correlation data. In this respect, it is useful to regard the $n$-point autoencoder as simulating a statistical inference experiment. The latent layer, consisting of $n$-point correlation functions, should be interpreted as a set of (possibly) sufficient statistics characterizing the system of interest. The encoder function summarizes complicated data by reconstituting the information presented therein in terms of the sufficient statistics, while the decoder leverages said statistics in order to predict out of sample. 
In light of the discussion provided in Section \ref{sub:_the_hamburger_problem}, one may also wish to regard the autoencoder as solving the restricted moment/cumulant reconstruction problem. In this sense, the $n$-point autoencoder presents the further benefit of quantifying the explicit information contained in subsets of correlation functions.

\subsubsection{Sampling Scheme}%
\label{ssub:importance_sampling}


For practical datasets such as those considered in the machine learning and data science communities considering all $n$-point functions for $n\geq3$ is both wasteful, and often computationally intractable. As such, an important consideration which goes into \eqref{eqn:cumulant_autoencoder_spaces_unmasked} is the choice of index set $\tilde{\mathcal{I}}$. A good index set should systematically remove potentially redundant or non-informative $n$-point functions to improve compute, while retaining enough correlators to adequately capture information about the underlying distribution. In other words, the construction of the space ${\mathcal{K}}^{\tilde{\mathcal I}}$ requires the introduction of an efficient sampling algorithm for $n$-point functions.

In this section we introduce such an algorithm for identifying a minimal set of $n$-point functions to sample to make the $n$-point autoencoder outlined in Section \ref{sub:ncoder} feasible to train for image data without sacrificing too much relevant information. In what follows the reader may benefit from having a definite example in mind. Recall that the MNIST dataset consists of approximately 60,000 training samples and 10,000 testing samples of handwritten digits 0 through 9. MNIST images have dimension $ L_x \times L_y = 28 \times 28$ and consist of a single color channel ($L_c = 1$); whilst MNIST images do have a (redundant) RGB representation, a common (and indeed the most computationally simple) approach is to consider a grayscale representation whereby each pixel value is given a value in the range $0 \to 255$ -- requiring 8 bits of information per site. Each image in the MNIST dataset has $28^2 = 784$ pixel locations (lattice sites), which can be flattened and described by a vector, provided the original relative position in 2D space is retained -- one can, of course, always reshape a vector $v_i$ back to a $28\times 28$ pixel matrix $m_{ij}$ with the simple transform $v_{28i + j} \leftrightarrow m_{ij}$.

Let $T$ be a set of arbitrarily ordered image samples. The first step we will take to make the data more manageable is to train the network described in Section \ref{sub:ncoder} in \emph{batches}. Without loss of generality, $T$ can be decomposed into a (disjoint) union of batches $B^\beta$ such that
\begin{equation}
	T = \bigcup_\beta B^\beta.
\end{equation}
Each flattened batch vector $(B^\beta)_r$ contains sampled images presented as vectors $y_i^j$ such that \eqref{eqn:flattened_training_data} becomes
\begin{equation}
	{B}^\beta = (B^{\beta; 0}_i, B^{\beta; 0}_{i+1}, \ldots, B^{\beta; j}_{|L|}, \ldots, B^{\beta; N_B}_{|L|}).
\end{equation}
Here, we have introduced the following notation: $B_i^{\beta; j}$ refers to the realized pixel output at the $i^{th}$ lattice site associated with the $j^{th}$ sample in the $\beta^{th}$ batch. Each batch consists of $N_B = |B|$ samples with $|B|$ chosen or $T$ slightly truncated such that $|T| \text{ mod } |B| = 0$. By construction, each batch contains unique samples, i.e.
\begin{equation}
	B^{\beta_1} \cap B^{\beta_2} = \emptyset \;  \forall \;  B^{\beta_1} \neq B^{\beta_2} \in T.
\end{equation}



Due to the typical size of image lattices being very large, the computation of $n$-point products over all lattice site combinations is computationally intensive. For example, in MNIST each image has
\begin{equation}
	\binom{28^2 + 3 - 1}{3} \approx 8 \times 10^7
\end{equation}
3-point combinations -- it is infeasible to consider them all for each batch. To overcome this difficulty, we need an approach which approximates the distribution of a carefully sampled subset of combinations. This subspace should be associated with an intentionally biased subset of points prioritizing the inclusion of points that concentrate around the subject of the image. As a result, highly-degenerate data is limited, reducing the computational burden and better encoding the short-range information implicit in the images. 

\begin{figure}[htpb]
	\centering
	\includegraphics[width=0.8\linewidth]{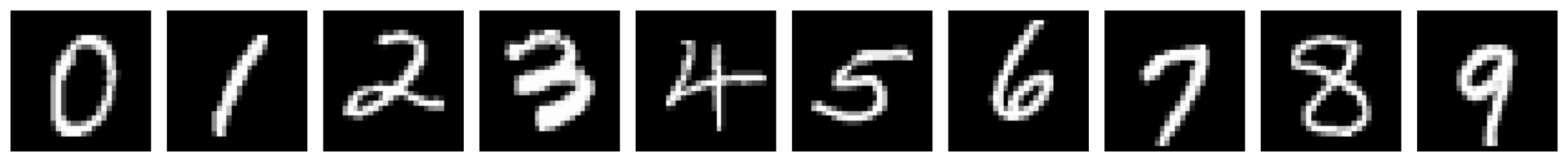}
	\caption{An example of each class of digit in the MNIST dataset.}%
	\label{fig:MNIST_example_digits}
\end{figure}

To motivate the approach, let us again consider an MNIST sample -- an example of each digit is included in Figure \ref{fig:MNIST_example_digits}.
The most identifying information is encoded in the white part of the image\footnote{By default, PyTorch assumes a normalized pixel value of 1 corresponds to white, so the digit appears as white on a black background. Some other frameworks may display the digit as written in black on a white background, with black being represented by $100\%$ pixel intensity.}, as opposed to the trivial background. As a result, amongst the entire space of 3-point cumulants only approximately 3\% of elements are non-zero. Sampling pixel combinations from a uniform distribution would therefore result in an extremely inefficient encoding; reconstruction would require prohibitively many correlators than are strictly necessary!

To overcome this difficulty, we employ a more nuanced sampling scheme by initially calculating a full set of arithmetic $n$-point cumulants for a small number of representative batches. By choosing an initialization/preprocessing regime such that black (or equivalently any highly degenerate color) is 0, like the PyTorch default, it is clear by inspection that the cumulants of these quantities will approximately vanish provided the image samples are sufficiently similar in that region (this is clear by inspection of the moment-cumulant formula). Samples are grouped such that the associated labels match -- the process assumes the samples are broadly similar and the constituent functions $\Phi(x)$ are independent and identically distributed for each $x$, and the procedure produces the best results for samples possessing only a single channel.

Let $T_\text{init}$ be a small subset of $N_\text{init}$ training batches from $T$
\begin{equation}
	T_\text{init} \subset T.
\end{equation}
Since $T_\text{init}$ is small, for sensible $n$ it is possible to compute the entire set of correlators for $N_\text{init}$ batches. We then introduce the following algorithm for selecting $n$-point functions to include in the latent space. 
First, one constructs a \emph{masked index set} $\tilde{\mathcal I}$ by identifying the lattice point combinations on which the sampled $n$-point functions from the initialization data set are nonzero: 
\begin{align} 
	\label{eqn:mask}
  \tilde{\mathcal I} := \big \{& I \mid \kappa_{\text{init}, \gamma}^I \neq 0 \, \forall \, I \in \mathcal I \text{ and } \forall \, \gamma \in \{0, \ldots, N_\text{init}-1\} \big\}.
\end{align}
Here, $\kappa^I_{\text{init}, \gamma}$ is the set of numerical $n$-point functions evaluated over batch $\gamma$ in the initialization set, obtained from taking the arithmetic mean over products of pixel site realizations in batch $\gamma$, as opposed to the entire dataset in \eqref{eqn:moment_of_full_dataset}.\footnote{It is reasonable to keep $N_\text{init}$ small provided the images are true IID samples. It may be the case, particularly if $N_\text{init}$ is too small, that $\mathcal K^{\tmi}$ is still too large to be computationally feasible -- if this occurs, one may consider a subset of the elements of $\mathcal{\tilde{I}}$ sampled from a uniform distribution without replacement.} It is useful to organize the index set as a collection of subsets coinciding with different order $n$-point functions, i.e.
\begin{equation} \label{ordered mask}
	\tilde{\mathcal I} \equiv \{ \tilde{\mathcal I}^{k_i}, \ldots, \tilde{\mathcal I}^{k_f} \}.
\end{equation}
In other words, $ \tilde{\mathcal I}^n$ is the set of all $n$-point combinations of lattice points with nonzero $n$-point functions over all batches in the initialization set.

In principle we could simply use all of the points selected in \eqref{ordered mask} as an informed sample of the relevant $n$-point functions. However, it still may be the case that the set of points implicated by \eqref{ordered mask} is too large to be computationally feasible, especially for higher $n$-point functions. To remedy this issue we further introduce the $n$-tuple $(\alpha_{k_i}, \ldots, \alpha_{k_f})$ which denote `acceptance parameters' quantifying the proportion of persistently non-zero cumulant combinations of each order to be included in the model. Given these acceptance parameters we simply select a uniform random sample of the lattice point combinations included in \eqref{ordered mask} with given acceptance parameters for each $n$. For simplicity, we refer to the reduced index set also as $\tilde{\mathcal{I}}$. The index set $\tilde{\mathcal I}$ defines a reduced set of $n$-point correlation functions deemed relevant according to the initialization sample. 


\begin{figure}[htpb]
    \centering
    \begin{tikzpicture}[scale=0.75,transform shape,->,>=stealth]

        \node[input] (input) {Batched samples};
        \node[hidden, below=of input] (encoder_hidden) {Encoder};
        \node[latent, below=of encoder_hidden] (code) {Masked $n$-pt functions};

        \node[hidden, below=of code] (decoder_hidden) {Decoder};
        \node[output, below=of decoder_hidden] (output) {Representative sample};

        \node[redbox, fit=(input) (encoder_hidden) (code), label={above:\textit{Initial train}}] {};

        \node[bluebox, fit=(code) (decoder_hidden) (output), label={below:\textit{Second train}}] {};

        \draw[->] (input) -- (encoder_hidden);
        \draw[->] (encoder_hidden) -- (code);
        \draw[->] (code) -- (decoder_hidden);
        \draw[->] (decoder_hidden) -- (output);

    \end{tikzpicture}
    \caption{Brief schematic overview of the `NCoder' setup.}%
    \label{fig:ncoder_setup}
\end{figure}

\section{Experimental implementatation}%
\label{sec:experimental_setup}

In prior sections, we presented a theoretical method for encoding the distribution of a system using numerically evaluated $n$-point functions drawn from a sample of realizations of some underlying data-generating distribution. In this section, we outline a practical numerical method for performing said encoding/decoding by means of a pair of multi-layer perceptron networks.

An underlying assumption of this exercise is that the space of n-point functions constitutes a sufficient description of the samples. That is, for some selection of masked $n$-point functions as in \eqref{eqn:cumulant_autoencoder_spaces_unmasked},
\begin{equation}
	\label{eqn:cumulants_collection}
K_{\text{eff}} = K^{\tilde{\mathcal I}},
\end{equation}
the distribution of the training samples may be resolved up to a chosen level of precision, $\epsilon$.\footnote{This idea is defined more concretely in Appendix \ref{sub:cumulant_space_equivalence_classes}.} Due to the vast combinatoric factors as $n$ grows larger than 3, it is only computationally feasible to probe $n$-point functions up to the third order. Nevertheless, in theory, as has been argued throughout this work, increasing the order of the $n$-point functions which are included should yield better reconstructions of the data. Our analysis examines the practical quantitative effect of adding higher $n$-point functions to reconstructions by  beginning with latent layers which involve only $1$-point data and sequentially incorporating $2$- and $3$-point functions, ensuring that the computational load is managed via the selection of a suitable masking as presented in \ref{ssub:importance_sampling}. 


\subsection{Encoder setup}%
\label{sub:encoder_setup}
In order to examine how data is retained within $n$-point functions, we first construct a neural network which estimates the distribution of $n$-point functions of a dataset given batches of realized input samples.

\paragraph{Encoder architecture}%
\label{par:encoder_architecture}
The encoder model is a simple feed-forward fully-connected architecture, composed of alternating ReLU activation functions and a pair of linear (dense) layers with trained weights and fixed trivial biases. The encoder architecture is described as
\begin{equation}
	\chi^{N_B} \overset{f_0}{\mapsto} H_0 \overset{f_1}{\mapsto} H_1 \overset{f_2}{\mapsto} \eta,
\end{equation}
where $\chi$ is the message space, $H_l$ the hidden layers, and $\eta$ the latent space. The message space $\chi^{N_B}$ is identified with the batch $B^\beta$, and the latent space $\eta \subset R^n$ (where $n$ is the size of the masked cumulant vector $n := |\tilde{\mathcal I}|$).

The maps $f_l$ between layers $L_l$ are defined through
\begin{equation}
	(L_{l+1})_i := f_l (B^\beta_i) = \sigma_l ((W_l)_{ij} (L_l)_j + (b_l)_i),
\end{equation}
where $W_l, b_l$ are the \textit{weights matrices} and \textit{bias vectors} respectively, $L_l = L_l(B^\beta)$, and the activation functions $\sigma_l(B)$ are set to be the \textit{Rectified Linear Unit (ReLU)} defined through
\begin{equation}
	\sigma_l (B_i) = \text{ReLU}(B_i) := 
	\begin{cases} B_i \quad \text{if} \quad B_i>0,\\ 
		0 \quad \text{otherwise}
	\end{cases}
\end{equation}

To simplify computations, we consider the network with bias vectors $\vec (b_l)_i = 0$. The ReLU activation is arguably one of the simplest activation functions one may choose, however it is subject to the so-called `vanishing gradients' problem: since the function is entirely constant for negative data, the gradient of the activation function is 0 for all data; the neuron can become stuck in a `dead' state and goes to 0 throughout training. One alternative is the Leaky ReLU activation, which modifies the ReLU activation so that the original data is suppressed by a small positive constant multiple when the argument is negative. Heuristic exploration has shown the network training and validation losses are within a few percent independent of whether \textit{ReLU, LeakyReLU, or GELU} activation functions are chosen.

\paragraph{Encoder training}%
\label{par:encoder_training}
A simple loss function which yields good results is the \textit{mean-square error\footnote{The MSE loss is also the $L^2$ loss (down to some multiplicative constant).} (MSE)} defined by
\begin{equation}
	\label{eqn:mse_def}
  \mathcal L(K^{\tilde{\mathcal I},\beta},\, K_\text{pred}^{\tilde{\mathcal I},\beta}) = \frac{1}{N_B} \bigg| K^{\tilde{\mathcal{I}},\beta} - K_\text{pred}^{\tilde{\mathcal I},\beta}\bigg|^2.
\end{equation}
Here, $K_\text{pred}^{\tilde{\mathcal I},\beta}$ is the output of the encoder evaluated on the batch $\beta$, while $K^{\tilde{I},\beta}$ are the numerically evaluated cumulants from the data. This choice of loss, of course, is not-unique -- one could equivalently consider the \textit{Kullback-Leibler divergence} for example. An example of the training loss is given in Figure \ref{fig:1_mnist}.

\subsection{Decoder setup}%
\label{sub:decoders_setup}

The decoder network, whilst disparate, is taken to have a similar architecture to that of the encoder. Instead of mapping to a larger latent space, the decoder acts as a map from the space of $n$-point functions into the space of samples. In other words, the decoder attempts to reconstruct data which are consistent with the inserted summary statistics. As has been stressed in previous section, the message spaces of the encoder and decoder networks are different in our $N$-coder architecture.


\paragraph{Decoder training}%
\label{par:decoder_training}

Given an input vector $K^{\tmi, \beta}$, the decoder network produces an output vector $\zeta^{\beta}$ which is a collection of $m_\text{out}$ full samples reconstructed from the summary statistics implied by the batch $\beta$. The output vector, which is simply a collection of generated images, has associated with it a vector of masked $n$-point functions $\Gamma^{\tmi, \beta}$. These are simply the arithmetic $n$-point correlation functions of the reconstructed batch. Similarly to the encoder, the decoder is trained subject to the MSE loss 
\begin{equation}
    \mathcal L(\Gamma^{\tmi, \beta}, K^{\tmi, \beta}) := \frac{1}{N_B} \bigg | \Gamma^{\tmi, \beta} - K^{\tmi, \beta} \bigg |^2.
\end{equation}
More general loss functions may be employed.

\subsection{Output validation}%
\label{sub:validation}

During the training process, the performance of the network (loss) is measured by comparing the predicted and the sampled $n$-point functions evaluated over each batch. One expects the network to learn features of the training samples from the $n$-point functions and reproduce samples which are representative of such features. Nevertheless, we require a more refined measure for the effectiveness of the NCoder to validate its performance. Simply measuring how well the network estimates the true and approximate $n$-point functions, while a valid measure of the efficiency of the combined encoder and decoder, obscures the quality of the samples. 


A naive first idea for validation loss might be to compare the MSE or some equivalent pixel-to-pixel loss between the output of the encoder and each of the input samples e.g.
\begin{equation}
\label{eqn:mse_validation_loss} 
	\mathcal L = \sum_\beta^{N_B} \sum_j^{|B|} |{y}^\beta - {y}^\beta_{\text{pred}, j}|^2.
\end{equation}
However, \eqref{eqn:mse_validation_loss} also proves to be a poor proxy for the quality of the NCoder: as this loss is adversely affected by transformations (e.g. rotation, translation, etc) which lead to large losses even when outputs and samples are otherwise identical. Additionally, MSE is not robust against atypical samples. As our goal is to produce a network that trains on persistent features, it would not be prudent to penalize the NCoder for weighting samples with a lower realization probability comparatively lower than those representative of features which appear more frequently.

An alternative validation measure which is more evocative of our goal is given by the classification accuracy of the reconstructed sample. In other words, we measure the performance of the NCoder by testing whether the data it generates is correctly classified by an efficient classification algorithm appropriate to the given task. Classifiers, particularly those with convolutional layers on the input, are significantly less likely to be affected by transformations of the input, instead focusing on more holistic features. If the reconstructed data returns a high classification accuracy, we can at least validate that the NCoder is producing samples which can reliably be recognized as coming from the distribution we intended to reconstruct. For a discussion of the classifier network we are using in our validation loss we refer the reader to Appendix \ref{sec:classifiers}.


\subsection{MNIST reconstructions}%
\label{sub:mnist_reconstructions}

We are now prepared to apply our NCoder architecture to a specific data task, namely learning features of samples from the MNIST dataset. In this demonstration, we consider a subset of the MNIST dataset consisting of a single class of samples: those depicting the digit `7'\footnote{One is free, of course, to choose any class of sample from the dataset}. To be more concrete, the training batches $B^\beta$ are composed of drawn samples of sevens, each assumed to be IID. The network parameters are listed in Figure \ref{fig:encoder_decoder_parameters} in Appendix \ref{app: training}. Figure \ref{fig:mnist_grid} presents the training losses for the decoder of the neural networks encoding single point, single- and two-point functions, and single-, two-, and three-point functions. 
 
\begin{figure}[htpb]
	\centering
	\includegraphics[width=0.8\linewidth]{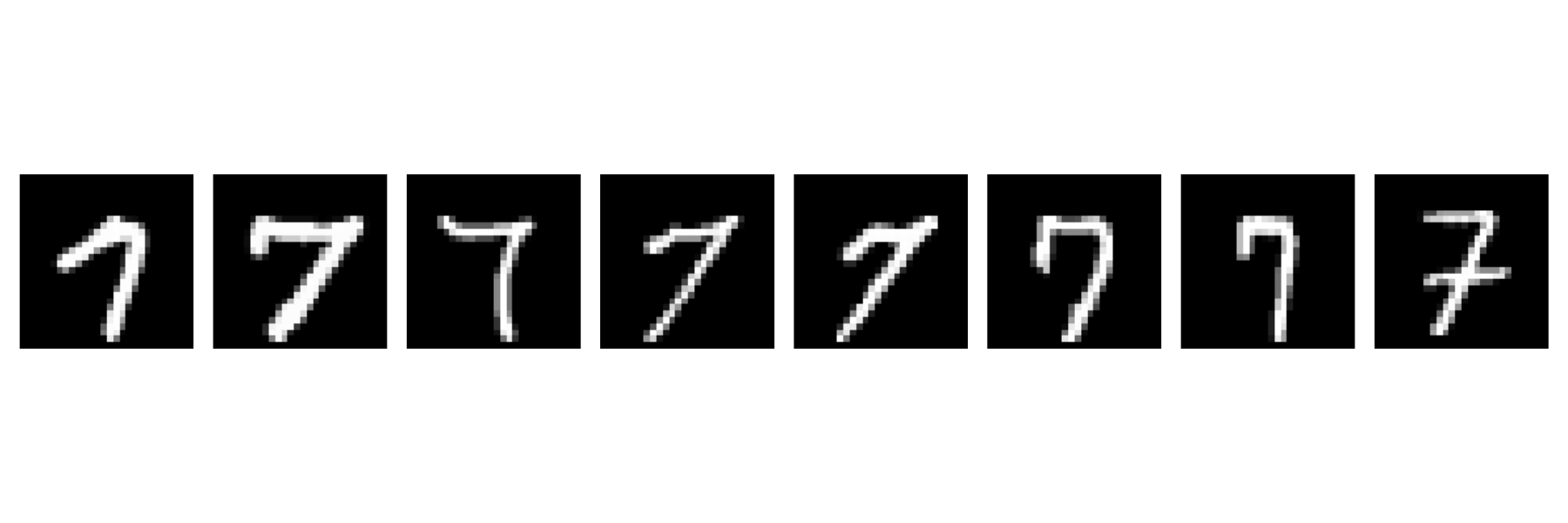}
	\caption{An example of each samples of the digit `7' in the MNIST dataset.}%
	\label{fig:MNIST_example_of_sevens}
\end{figure}

\begin{figure}[h!]
\centering

	\begin{subfigure}{0.3\textwidth}
		\includegraphics[width=\linewidth]{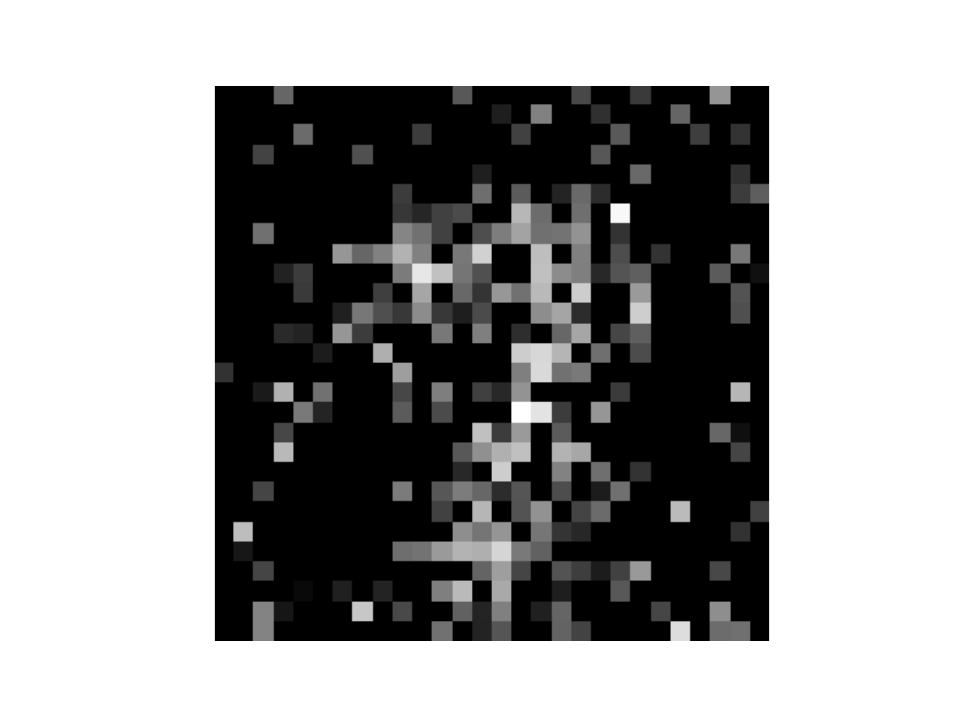}
		\caption{Reconstruction of an MNIST seven from single-point cumulants only.}%
		\label{fig:mnist_kikf1}
	\end{subfigure}\hfil
		\begin{subfigure}{0.3\textwidth}
		\includegraphics[width=\linewidth]{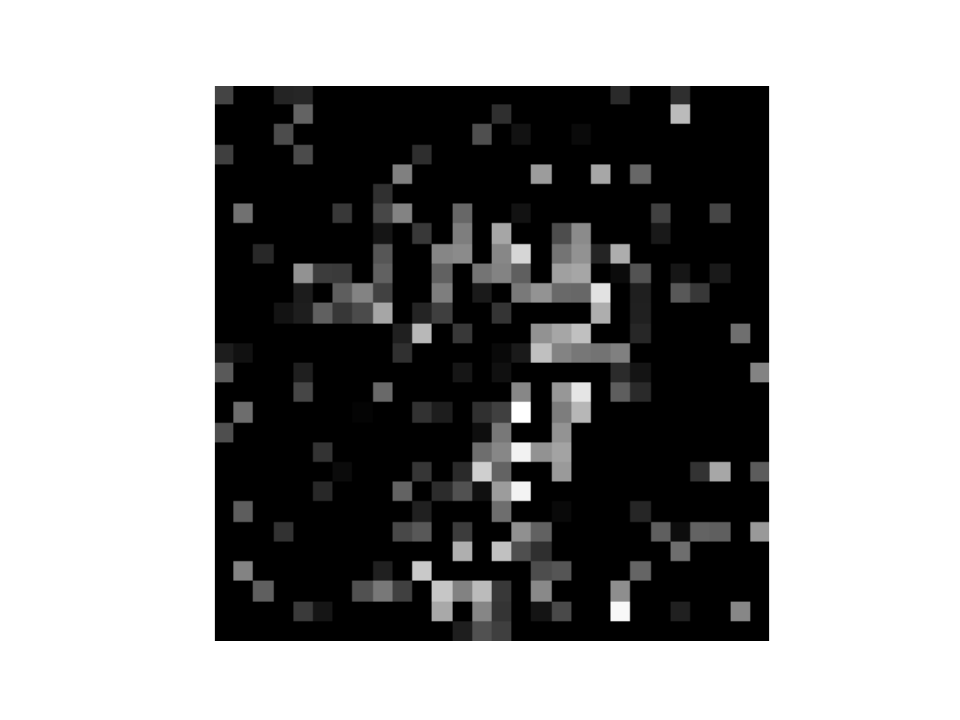}
		\caption{Reconstruction of an MNIST seven from 1- and 2-point cumulants.}%
		\label{fig:mnist_kikf12}
	\end{subfigure}\hfil
		\begin{subfigure}{0.3\textwidth}
		\includegraphics[width=\linewidth]{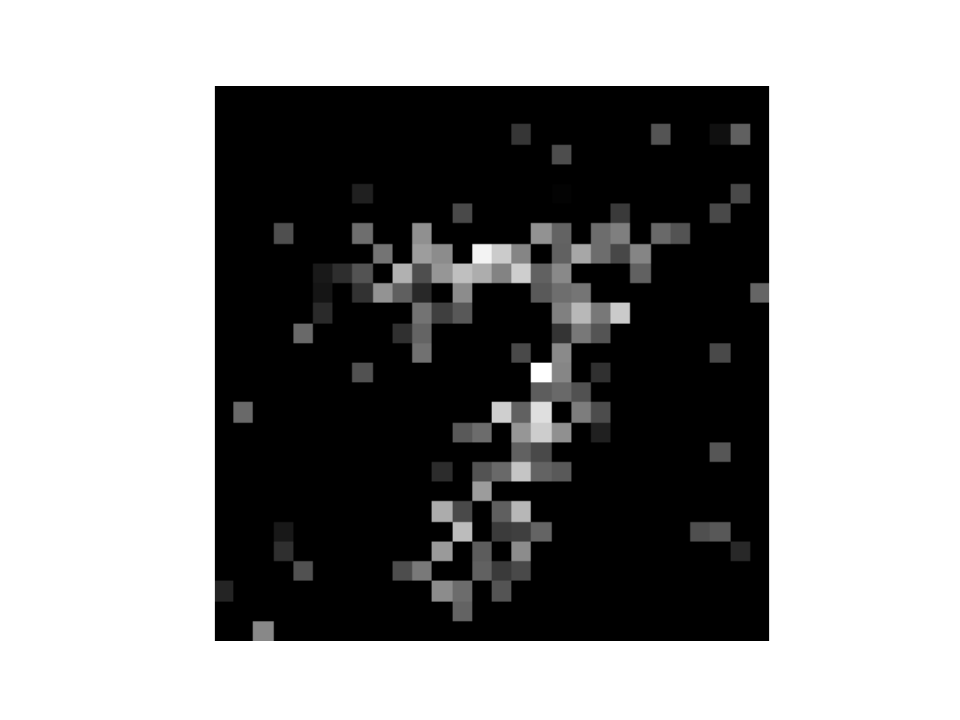}
		\caption{Reconstruction of an MNIST seven from 1-, 2-, and 3-point cumulants.}%
		\label{fig:mnist_kikf123}
	\end{subfigure}\hfil
	
	\begin{subfigure}{0.3\textwidth}
      \includegraphics[width=\linewidth]{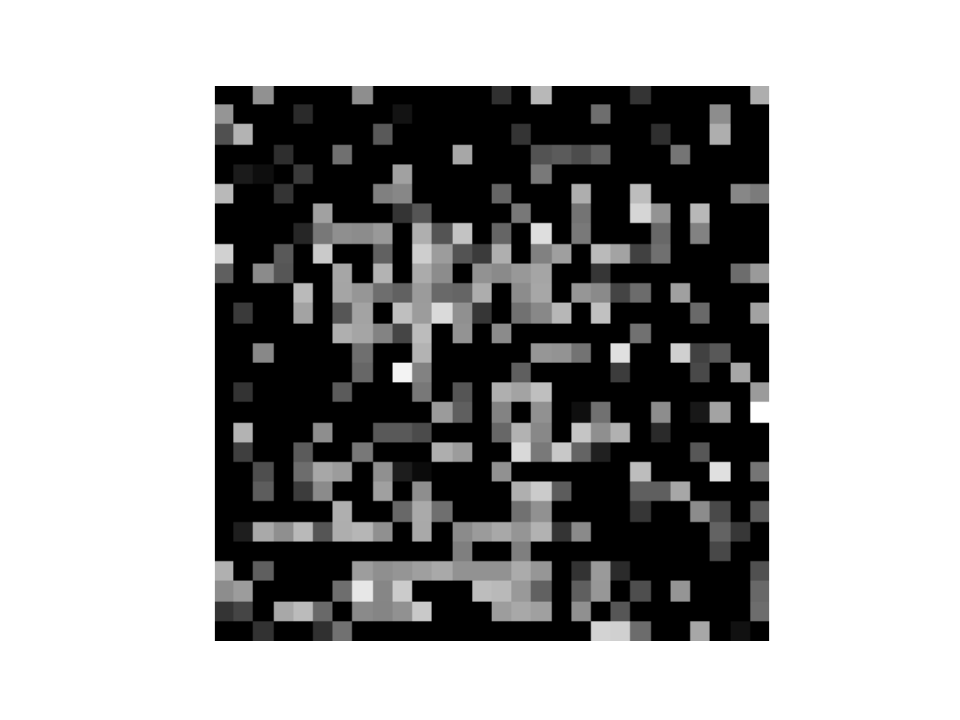}
		\caption{Reconstruction of an MNIST seven from single-point moments only.}%
		\label{fig:mnist_kikf1_moment}
	\end{subfigure}\hfil
	\begin{subfigure}{0.3\textwidth}
      \includegraphics[width=\linewidth]{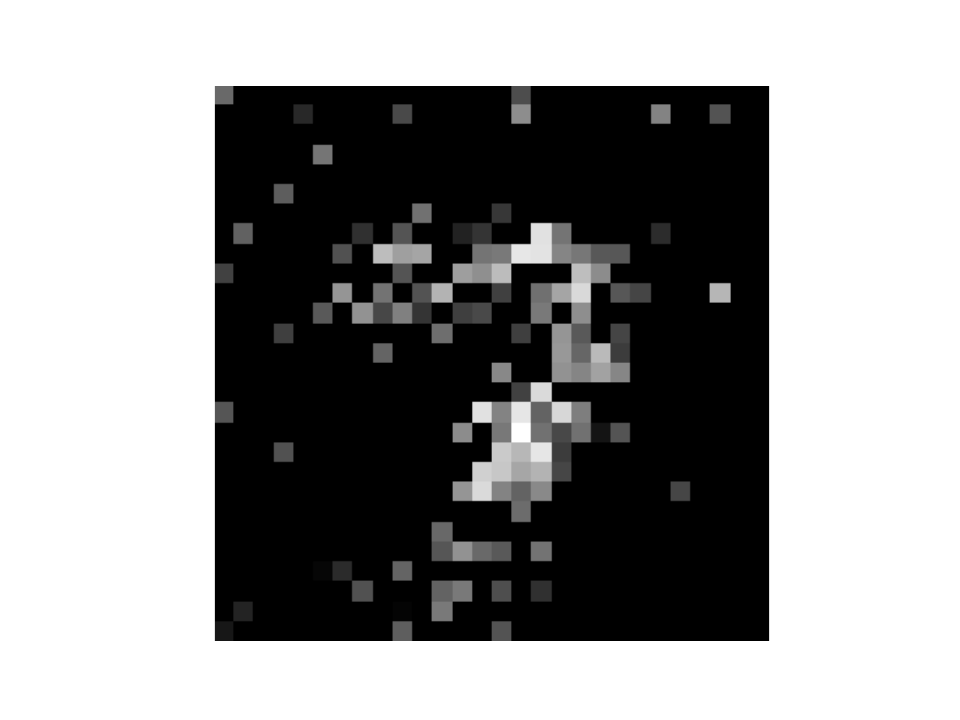}
		\caption{Reconstruction of an MNIST seven from 1- and 2-point moments.}%
		\label{fig:mnist_kikf12_moment}
	\end{subfigure}\hfil
	\begin{subfigure}{0.3\textwidth}
      \includegraphics[width=\linewidth]{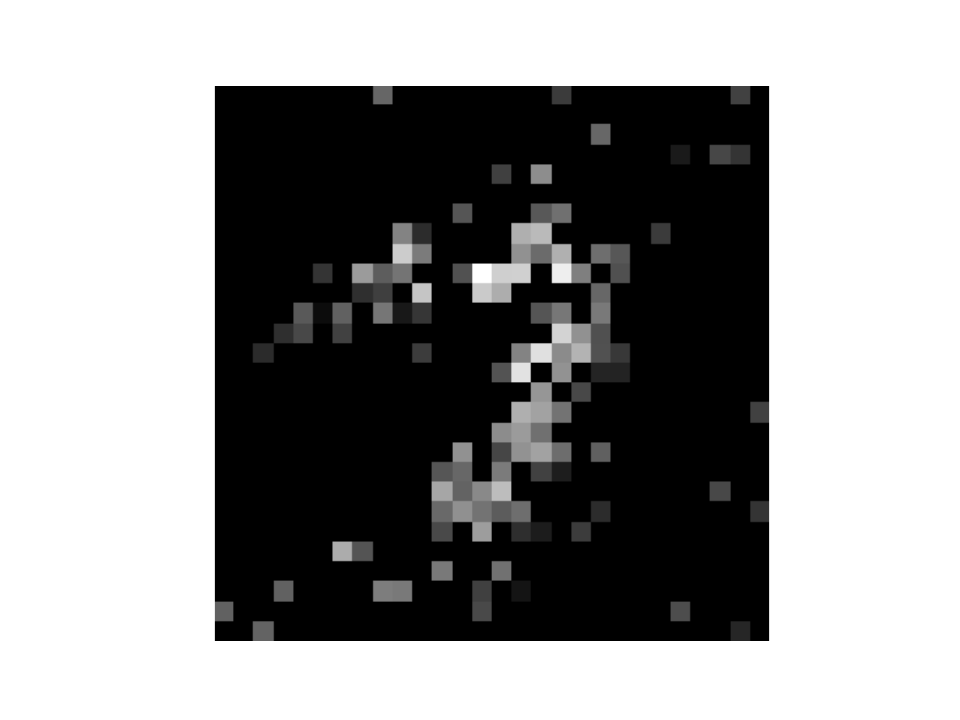}
		\caption{Reconstruction of an MNIST seven from 1-, 2-, and 3-point moments.}%
		\label{fig:mnist_kikf123_moment}
	\end{subfigure}
	\caption{Reconstructions of samples of sevens from the MNIST dataset with varying $n$-point functions included. Cumulant reconstructions presented on the top, moments on the bottom.}
	\label{fig:mnist_reconstruct}

\end{figure}

In Figure \ref{fig:mnist_reconstruct} the reader can find reconstructed MNIST images trained on samples of the number seven. For reference, in Figure \ref{fig:MNIST_example_of_sevens}, we include an example batch of samples depicting the digit `7' used for training. The reconstructed samples are organized from left to right by the highest order $n$-point correlation functions which are included in the latent layer of the NCoder network. By inspection, increasing the number of $n$-point functions improves the visual acuity of the reconstructed images. This is corroborated by the validation measure in terms of the classifier accuracy, as can be seen in Figure \ref{fig:val_accuracies}. Here, the validation accuracy is measured using a 2-way classifier technique described in detail in Appendix \ref{sec:classifiers}. The stated accuracy is an average over 10 separate training runs of the classifier with random training sample ordering to ensure numerical stability. 

\begin{figure}[htpb]
    \centering
    \begin{tabular}{|c|c|c|}
        \hline
                                             & \textbf{Accuracy/\% (cumulants)} & \textbf{Accuracy/\% (moments)} \\
        \hline
        \textit{Control: Original dataset}   & 99                               & 99\\
        1-, 2-, and 3-point functions        & 100                              & 74\\
        1- and 2-point functions             & 72                               & 53\\
        1-point only                         & 3                                & 0.3\\
        \hline
    \end{tabular}
    \caption{Validation accuracies of reconstructions with varying $n$-point functions.}%
    \label{fig:val_accuracies}
\end{figure}

\subsection{FashionMNIST reconstructions}%
\label{sub:fashionmnist}

\begin{figure}[h!]
\centering

	\begin{subfigure}{0.3\textwidth}
		\includegraphics[width=\linewidth]{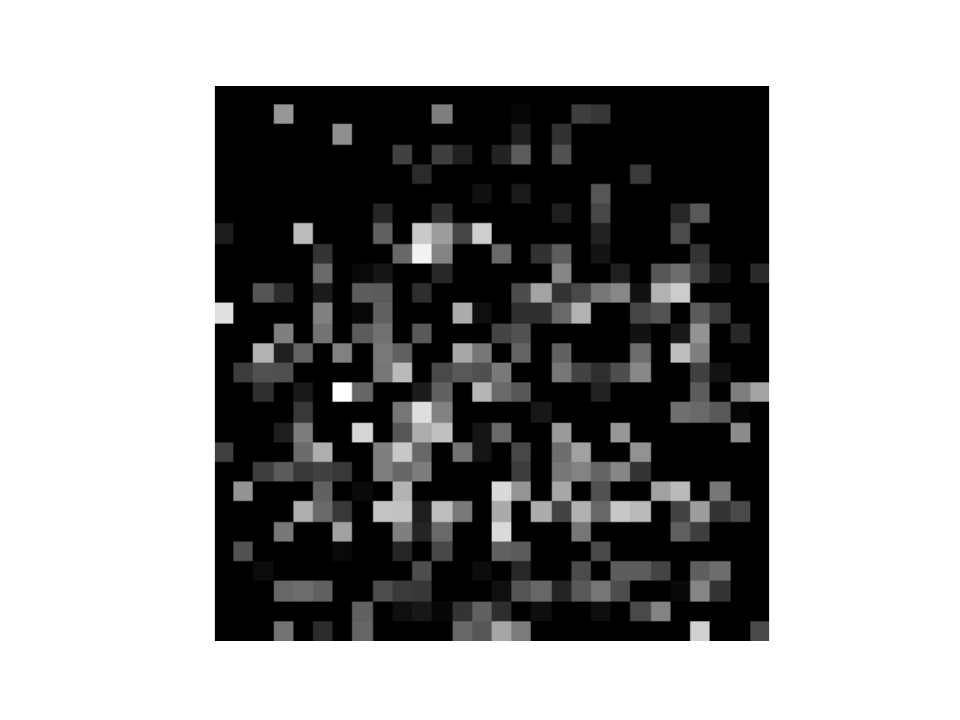}
		\caption{Reconstruction of a FashionMNIST bag from single-point cumulants only.}%
		\label{fig:mnist_kikf1}
	\end{subfigure}\hfil
	\begin{subfigure}{0.3\textwidth}
		\includegraphics[width=\linewidth]{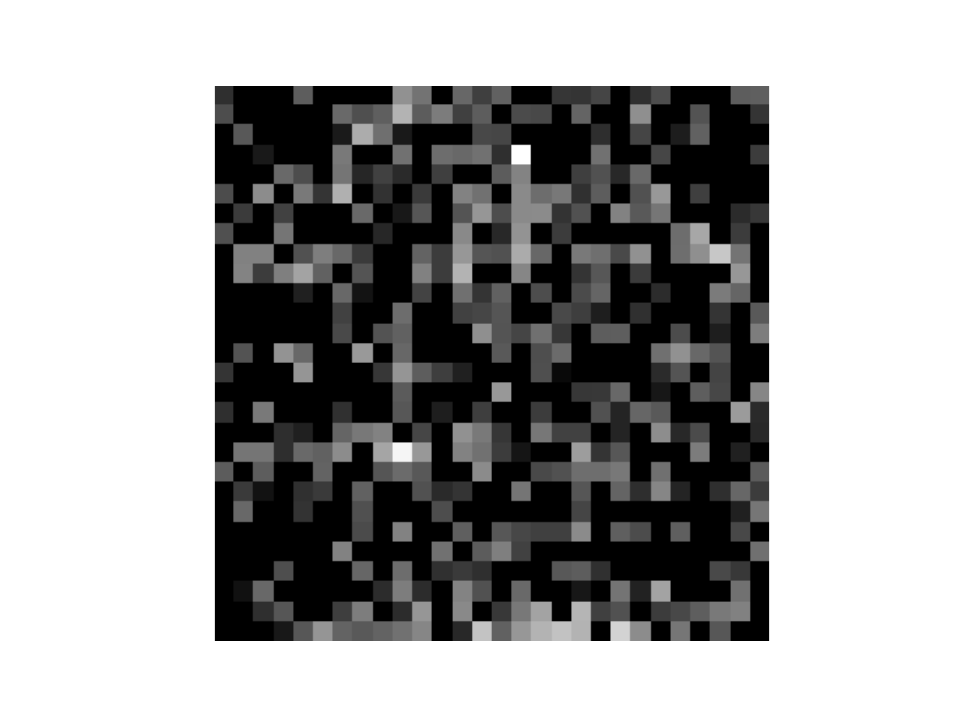}
		\caption{Reconstruction of a FashionMNIST bag from 1- and 2-point cumulants.}%
		\label{fig:mnist_kikf12}
	\end{subfigure}\hfil
  \begin{subfigure}{0.3\textwidth}
		\includegraphics[width=\linewidth]{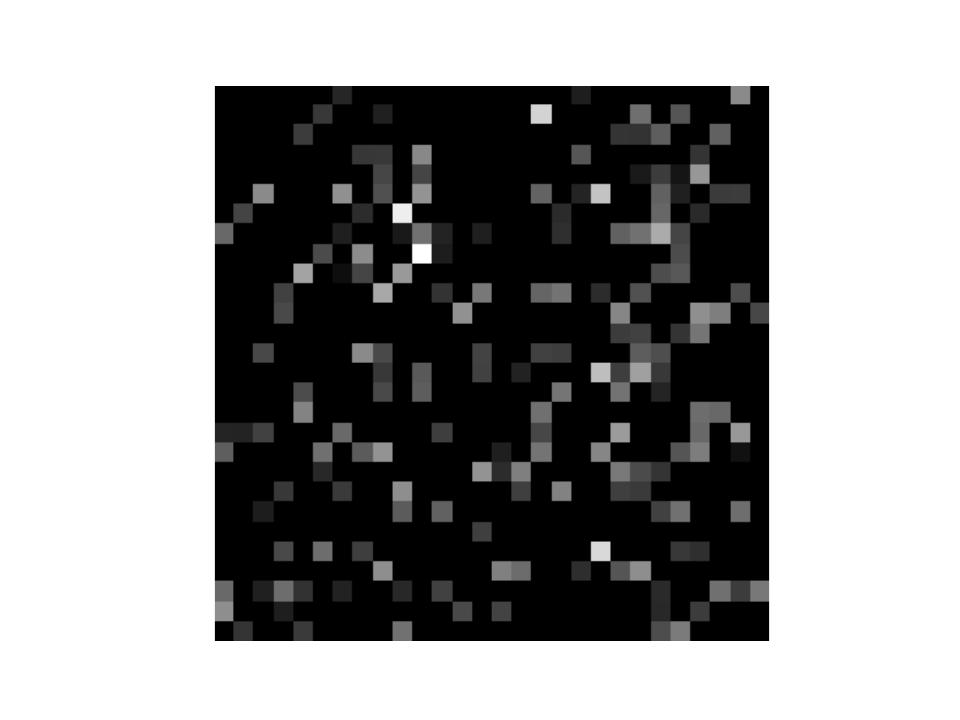}
		\caption{Reconstruction of a FashionMNIST bag from 1-, 2-, and 3-point cumulants.}%
		\label{fig:mnist_kikf123}
	\end{subfigure}\hfil

	\begin{subfigure}{0.3\textwidth}
      \includegraphics[width=\linewidth]{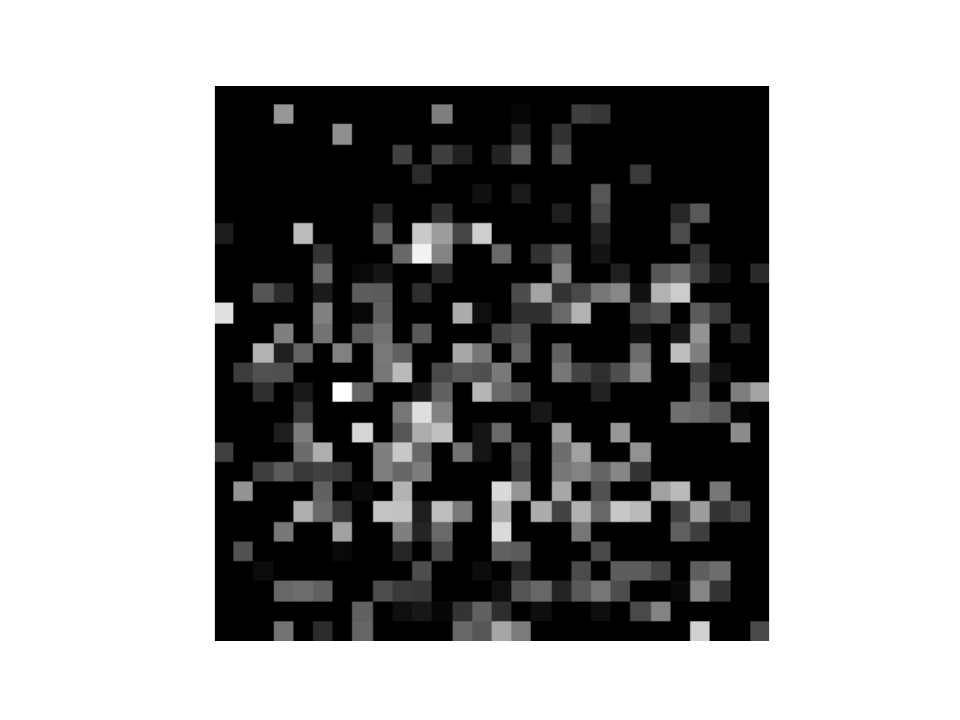}
		\caption{Reconstruction of a FashionMNIST bag from single-point moments only.}%
		\label{fig:mnist_kikf1_moment}
	\end{subfigure}\hfil
	\begin{subfigure}{0.3\textwidth}
      \includegraphics[width=\linewidth]{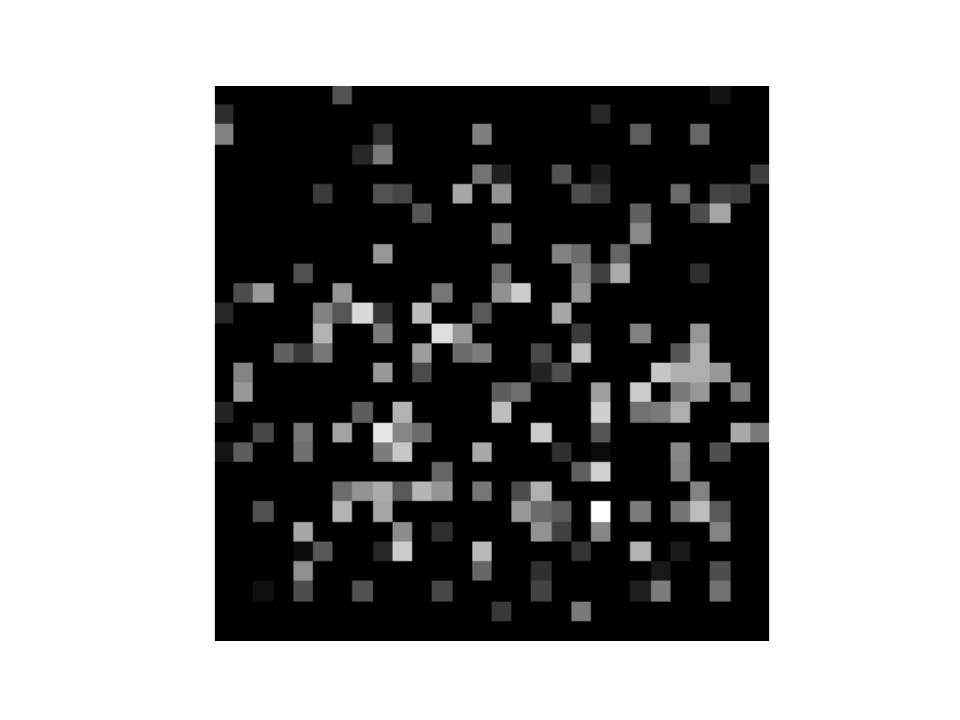}
		\caption{Reconstruction of a FashionMNIST bag from 1- and 2-point moments.}%
		\label{fig:mnist_kikf12_moment}
	\end{subfigure}\hfil
	\begin{subfigure}{0.3\textwidth}
      \includegraphics[width=\linewidth]{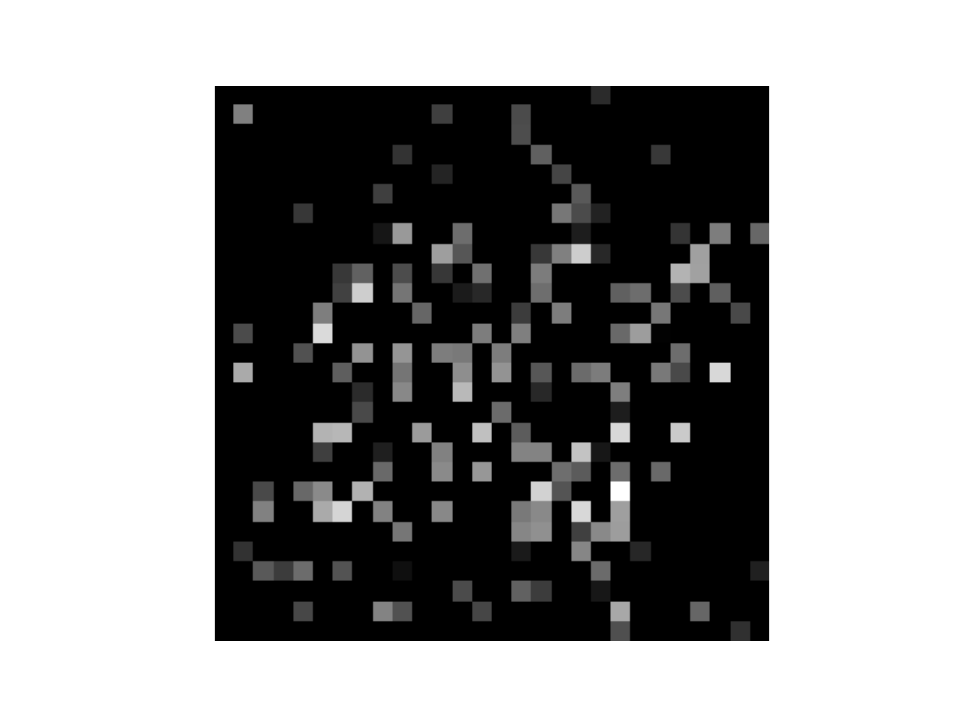}
		\caption{Reconstruction of a FashionMNIST bag from 1-, 2-, and 3-point moments.}%
		\label{fig:mnist_kikf123_moment}
	\end{subfigure}

	\caption{Reconstructions of samples of bags from the FashionMNIST dataset with varying $n$-point functions included. Cumulant reconstructions presented on the left, moments on the right.}
  \label{fig:fashion_mnist_reconstr}
\end{figure}

The success of the NCoder on MNIST data is encouraging, but it leaves open the question of whether this approach is adaptable to other, potentially more complicated distributions. As a simple first pass at this problem we have applied the same NCoder architecture to the problem of reconstructing samples from the Fashion MNIST data set. The performance is not nearly as good for this data set, and even with up to three points functions included in the latent layer the validation classifier cannot consistently identify reconstructed samples. Still, there is a sense in which reconstructed data does, at least visually, improve in acuity as higher correlation data is included in the latent layer -- see Figure \ref{fig:fashion_mnist_reconstr}.

\section{Predicting the efficacy of the NCoder approach} \label{sec: Ising Trial}

When comparing the MNIST and Fashion MNIST results, we observe a stark difference in the efficacy of the NCoder reconstruction approach. In this section, we aim to understand this difference. The NCoder architecture is built upon the assumption that there is a hierarchy of importance amongst $n$-point correlation functions. In the implementation, one is free to cut off the architecture at a chosen $n$-point function with the expectation that higher $n$-point functions are less relevant for controlling the behaviour of the system under study. In the experiments reported upon in this note, we have truncated the NCoder at the $3$-point function, while also prioritizing the significance of the $1$- and $2$-point functions by limiting the number of $3$-point instances that are incorporated into the training sample\footnote{In practice one must limit the number of $n-$ point functions for some value of $n$ as the number of samples to be included grows combinatorially.}. This can be viewed as an assumption that the underlying data which is being modelled is essentially Gaussian (e.g. governed primarily by the $2$-point function) but with some small non-Gaussianities which the NCoder learns through the $3$-point function. From this point of view, the difference in the quality of the output between MNIST and Fashion MNIST reveals a difference in the underlying statistics of these datasets e.g. MNIST is well modelled by a GRP whereas Fashion MNIST should not be truncated in this way. 

To test this hypothesis, we will now apply the NCoder architecture to modelling data in which we can control the `scale' of deformations to an otherwise Gaussian system. Taking further inspiration from theoretical physics, we consider datasets which are comprised of samples of lattice configurations from the next-nearest-neighbour (NNN) Ising model. The classical Ising model in 2D is a model often used in statistical physics as a playground for studying phase transitions and critical phenomena. Spins are arranged on a 2D lattice, where each site has an associated `spin' $\sigma_i = \pm 1$. In the presence of no external magnetic field, the model Hamiltonian is given by
\begin{equation}
H = -J_\text{NN} \sum_{\braket{i, j}} \sigma_i \sigma_j,
\end{equation}
where the notation $\braket{i,j}$ denotes a sum over all nearest neighbour pairs. It is typical to employ periodic boundary conditions such that each lattice site has the same number of neighbours. 

The NNN Ising model is an extension of the classical Ising model with an additional term in the Hamiltonian incorporating next-nearest neighbour interactions. For equally spaced square lattices, the NNN interactions include the spins associated to the four nearest  diagonal lattice sites (see Figure \ref{fig:schematic_of_nnn_ising}). For concreteness, the NNN Ising Hamiltonian is given by
\begin{equation}
H = -J_\text{NN} \sum_{\braket{i,j}} \sigma_i \sigma_j - J_\text{NNN} \sum_{\braket{\braket{i,j}}} \sigma_i \sigma_j,
\end{equation}
where $\braket{\braket{i,j}}$ represents all \textit{next-nearest-neighbour} interactions.
It is clear that sending $J_\text{NNN} \to 0$ recovers the nearest neighbour Ising model.

The NNN Ising model possesses two couplings, the `nearest neighbour' coupling and the `next-nearest neighbour coupling'. In the limit in which $J_{\text{NNN}} << J_{\text{NN}}$ we expect that the $2$-point function will be the dominant feature characterizing lattice configurations. As the next to nearest neighbor interaction grows in strength, higher point correlators will become comparable to the $2$-point function. So, if our hypothesis is correct, the efficacy of the NCoder model cutoff at the $2$-point function with minor corrections from a small number of $3$-point functions should diminish as a function of the coupling $J_{\text{NNN}}$. 

\begin{figure}
    \centering
    \includegraphics[width=0.5\linewidth]{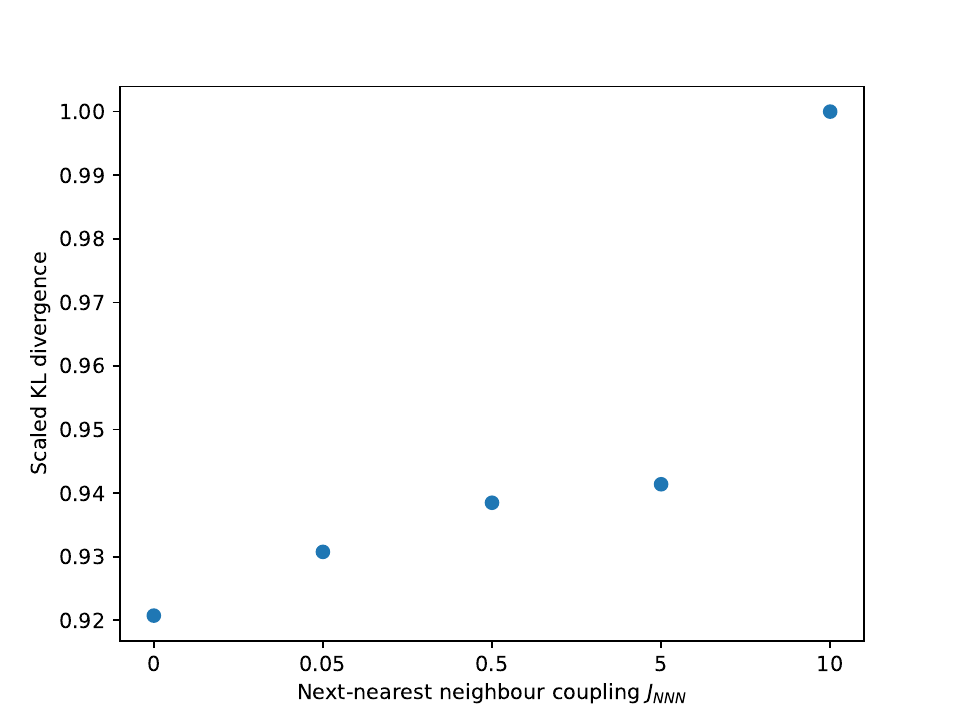}
    \caption{Relationship between KL divergence and the next nearest neighbour coupling $J_\text{NNN}$ for a set of 4000 samples over 5 averaged test networks.}
    \label{fig:nnn_kl}
\end{figure}

Using Monte Carlo sampling, we consider a dataset constituting lattice spin configurations for a fixed set of parameters with only the next-nearest neighbour coupling varied. We perform this analysis in the low-temperature regime such that the effective temperature $T$ of the system is sufficiently low $T$ ($T=2$) and the model is in the ferromagnetic phase. 
Figure \ref{fig:nnn_kl} shows the performance of the NCoder network as a function of increasing $J_\text{NNN}$. The vertical axis represents the KL divergence between input batches and output samples, averaged over 4000 samples. As expected, the KL-divergence increases monotonically. 

This study provides clarity to the usefulness of the NCoder architecture, but also an indication of its possible pitfalls. Constructing an NCoder which depends upon some number of correlation functions provides a way of probing the statistical structure of a given sample. If the reconstruction is efficacious with some chosen cutoff -- say up to $n^*$-point functions -- we can argue that the data has an effective model in terms of $n$-point functions with $n \leq n^*$. On the other hand, however, if a dataset is not well approximated by a model which includes contributions from lower $n$-point functions we are forced to incorporate more and more $n$-point data into our reconstruction algorithm. Although including higher $n$-point functions is possible in principle, it increases the computational demand of the architecture. For example, even to implement the current instance of the NCoder which includes only a small number of $3$-point functions is computationally costly. Nevertheless, the NCoder remains a useful diagnostic tool (e.g. for determining the order or relevant interactions for a given dataset) even in cases where it is not computationally feasible to use it as a tool for full reconstruction.

\begin{figure}
    \centering
\begin{tikzpicture}[scale=1.0]

\foreach \i in {0,...,4} {
    \foreach \j in {0,...,4} {

        \colorlet{spincolor}{black}

        \ifnum\i=1
            \ifnum\j=1
                \colorlet{spincolor}{red}
            \else\ifnum\j=3
                \colorlet{spincolor}{red}
            \fi\fi
        \fi
        
        \ifnum\i=3
            \ifnum\j=3
                \colorlet{spincolor}{red}
            \else\ifnum\j=1
                \colorlet{spincolor}{red}
            \fi\fi
        \fi
        
        \ifnum\i=1
            \ifnum\j=2
                \colorlet{spincolor}{blue}
            \fi
        \fi
        \ifnum\i=3
            \ifnum\j=2
                \colorlet{spincolor}{blue}
            \fi
        \fi

        \ifnum\i=1
            \ifnum\j=2
                \colorlet{spincolor}{blue}
            \fi
        \fi

        \ifnum\i=2
            \ifnum\j=3
                \colorlet{spincolor}{blue}
            \fi
        \fi
        
        \ifnum\i=2
            \ifnum\j=1
                \colorlet{spincolor}{blue}
            \fi
        \fi
        
        \filldraw[black] (\i,\j) circle (0.00); 

        \pgfmathtruncatemacro{\parity}{mod(\i+\j,2)}
        \ifnum\parity=0
            \draw[->, ultra thick, spincolor] (\i,\j-0.15) -- (\i,\j+0.3); 
        \else
            \draw[->, ultra thick, spincolor] (\i,\j+0.15) -- (\i,\j-0.3); 
        \fi
    }
}

\end{tikzpicture}
    \caption{Schematic of the nearest neighbouring spins (blue) and next-nearest neighbouring spins (red) for an arbitrary lattice site.}
    \label{fig:schematic_of_nnn_ising}
\end{figure}

\section{Discussion}%
\label{sec:discussion}

In this paper we introduced the NCoder, an autoencoder built around the concept of encoding image data in $n$-point correlation functions. The motivation behind the NCoder melds together intuition from QFT and data science where correlators are frequently used as a general class of statistics for summarizing data. In Section \ref{sub:_the_hamburger_problem} we formalized the role of these correlation functions both in terms of the moment/cumulant reconstruction problem and its perturbative solution via the Edgeworth expansion. The Edgeworth expansion, in particular, demonstrates how a full probability density can be reconstructed in terms of its complete set of correlation data. In this sense, the set of all correlators are sufficient for the task of performing inference on any class of probability models. 

The NCoder bears a strong resemblance to standard autoencoder architectures, but with the important caveat that the latent layer is not freely identified through the simultaneous training of the encoder and decoder. Rather, it is fixed and equal to the space of a prescribed set of $n$-point correlators. To accommodate this change, the NCoder trains its encoder and decoder functions separately. Of course, fixing the latent layer constrains the training and thereby results in lower performance than an unconstrained autoencoder\footnote{Indeed, the $n$-point function latent space is always an option for an unconstrained autoencoder, and thus the NCoder can only equal but never exceed the performance of an unconstrained autoencoder.}. However, the advantage of the NCoder architecture is that it opens up the hood of the neural network. As we described in Section \ref{sub:ncoder}, the NCoder can be interpreted as a hybrid between deep and conventional statistical learning. Although the encoder and decoder are still multi-layer perceptron neural networks, the architecture of the NCoder essentially asks the neural network to engage in a form of statistical inference by first learning summary statistics (the $n$-point functions) and then generating out-of-sample data from these summary statistics.

In principle, the NCoder should be capable of generating arbitrarily good samples when provided access to sufficiently many $n$-point functions. In practice, however, obtaining $n$-point functions is computationally intensive due to the large size of the images and the combinatorics involved. One of the main contributions of this note is a sampling algorithm, introduced in Section \ref{ssub:importance_sampling}, which identifies a subset of $n$-point functions which contain relevant information towards improving feature learning and image generation. When applied to sparse images, like MNIST numbers, which involve a relatively small number of relevant (e.g. nonzero) $n$-point correlators, the NCoder performs extremely well in generating recognizable data. A further benefit of the NCoder architecture is that it allows for the experimenter to access the importance of including higher $n$-point functions by changing the correlators which are included in the latent layer. For example, in Figure \ref{fig:mnist_reconstruct} it is clear how adding higher $n$-point functions successfully resolves finer detail in the image. 

Further experimenting with the NCoder for more detailed images however, like those included in the Fashion MNIST data set, has revealed some potential pitfalls. It is not possible to reconstruct Fashion MNIST images using an encoded layer comprised of a computationally feasible number of correlation functions. Nevertheless, there are important lessons to be learned from this state of affairs. The Fashion samples are far more structured and nuanced than the MNIST samples, so it is not surprising that higher $n$-point functions (that is beyond $n = 3$) are likely necessary to adequately resolve sufficient features of the underling distribution. From a physics perspective this echoes the question of whether perturbation series can be truncated at a particular finite order in the $n$-point correlation functions computed. For example, we might be inclined to say that MNIST is perturbatively complete at $3$-point functions (at least for the classification task we have presented), while Fashion MNIST is not. From a statistical inference perspective this is evocative of the question of sufficient statistics. That is, $1$-, $2$-, and $3$-point functions constitute a set of sufficient statistics for MNIST classification, but not for Fashion. The relationship between sufficiency and perturbative renormalizability is one that we hope to address in future work.

In a similar vein one can consider a generalization of the NCoder architecture in which, rather than prescribing the latent layer to consist of $n$-point correlation functions, one takes the latent layer to consist of a series of user defined `sufficient statistics'. This approach bears a close resemblance to recent work \cite{Faucett:2020vbu} in which the authors attempt to interpret a classifier neural network by iteratively constructing a decision function built only from interpretable statistics of the data which best approximates the decision rule of the network. The modified NCoder could hypothetically be used to identify the most important `human readable' statistics by first training both the encoder and decoder, and then running the encoder on its own to see which statistics it tends to single out as significant. An advantage of this approach is that the user defined functions in the new latent layer can be as computationally simple as the user would like, thereby overcoming the computational challenge of obtaining $n$-point correlators from the data. However, this is a dual edged sword as there are no guarantees the user defined functions will be sufficient. In forthcoming work we plan to apply this proposed modification to the NCoder to event classification tasks at the LHC. This is a data task which is particularly well suited to the NCoder approach since there exists a complete linear basis of interpretable statistics which span the space of all (IR safe) observables \cite{Komiske_2018,Collado_2021} thereby overcoming the aforementioned pitfall.

It is especially intriguing to consider how this generalization of the NCoder could work in conjunction with the Bayesian Renormalization scheme introduced in \cite{Berman:2022uov,Berman:2023rqb}. As was alluded to in the introduction, in \cite{Berman:2022uov,Berman:2023rqb} an information theoretic approach to the renormalization group was developed which assesses the relevance of model parameters according to their sensitivity to the acquisition of new data. Given a threshold on the maximal accuracy that a model can achieve (for example in terms of the quantity or quality of observed data) the information theoretic, or Bayesian, renormalization scheme discards parameters which do not affect the performance of the model up to the prescribed accuracy. In the language of data science, this entails an automated scheme for discarding so-called ``sloppy" parameters \cite{daniels2008sloppiness,Machta_2013,transtrum2015perspective,Abbott_2023}. From an information geometric perspective, these sloppy parameters constitute directions in model space that are highly compact, and therefore provide a direct analogy to the short distance interactions which are coarse grained in standard RG for local field theories. Performing a Bayesian renormalization of a `sufficient statistic' NCoder should therefore allow for an experimenter to explore how interpretable statistics `flow' under the decimation of network parameters. We anticipate that this approach may be used to explore the space of neural networks in a manner completely analogous to the exploration of the space of field theories which is facilitated by the standard renormalization group. 

\appendix

\section*{Acknowledgements}%
\label{sec:acknowledgements}

We wish to thank Jim Halverson and Anindita Maiti for introducing us to the Edgeworth expansion and explaining their work on the NN/FT correspondence at StringData 2023. We also wish to thank Jessica Howard for her suggestions about variations of the NCoder architecture and other particle physics oriented applications of autoencoders. Finally, we'd like to thank Edward Hirst for our many enlightening discussions, and insightful comments on the manuscript. DSB and AGS acknowledge support from Pierre Andurand over the course of this research, and DSB is also supported by the Science and Technology Facilities Council (STFC) Consolidated Grant ST/T000686/1 “Amplitudes, strings \& duality.”  MSK is supported through the Physics department at the University of Illinois at Urbana-Champaign.

\section*{Data and code availablity}%
\label{sec:data_and_code_availablity}

The NCoder code is available from GitHub at \url{https://github.com/xand-stapleton/ncoder}.

\section{Parameters and training losses for networks} \label{app: training}
Where stated, moving averages of window size 10 are used to smooth the noise in the training plots and improve readability.

\begin{figure}[htpb]
	\centering
	\begin{tabular}{cc}
		\textbf{Network parameter} & \textbf{Parameter value} \\
		\hline
    Learning rate (Enc)                 & $10^{-3}$ \\
    Learning rate (Dec)                 & $10^{-3}$ \\
    Optimizer                           & Adam\\
    Training epochs (Enc)               & $8$\\
    Training epochs (Dec)               & $65$\\
    Batch size                          & $16$ \\
		1-point acceptance param $\alpha_1$ & $1$\\
		2-point acceptance param $\alpha_2$ & $1$\\
    3-point acceptance param $\alpha_3$ & $1.7 \cdot 10^{-2}$
	\end{tabular}	
	\caption{Noteworthy NCoder parameters for the MNIST reconstructions (cumulants and moments).}%
  
	\label{fig:encoder_decoder_parameters}
\end{figure}

\begin{figure}[htpb]
	\centering
	\begin{tabular}{cc}
		\textbf{Network parameter} & \textbf{Parameter value} \\
		\hline
    Learning rate (Enc)                 & $10^{-3}$ \\
    Learning rate (Dec)                 & $10^{-3}$ \\
    Optimizer                           & Adam\\
    Training epochs (Enc)               & $8$\\
    Training epochs (Dec)               & $65$\\
    Batch size                          & $16$ \\
		1-point acceptance param $\alpha_1$ & $1$\\
		2-point acceptance param $\alpha_2$ & $0.8$\\
    3-point acceptance param $\alpha_3$ & $10^{-3}$
	\end{tabular}	
	\caption{Noteworthy parameters for the encoder and decoder networks on FashionMNIST (cumulants and moments).}%
	\label{fig:encoder_decoder_parameters}
\end{figure}

\newpage

\begin{figure}[ht!]
    \centering
    
    \begin{subfigure}[b]{0.4\textwidth}
        \centering
        \includegraphics[width=\textwidth]{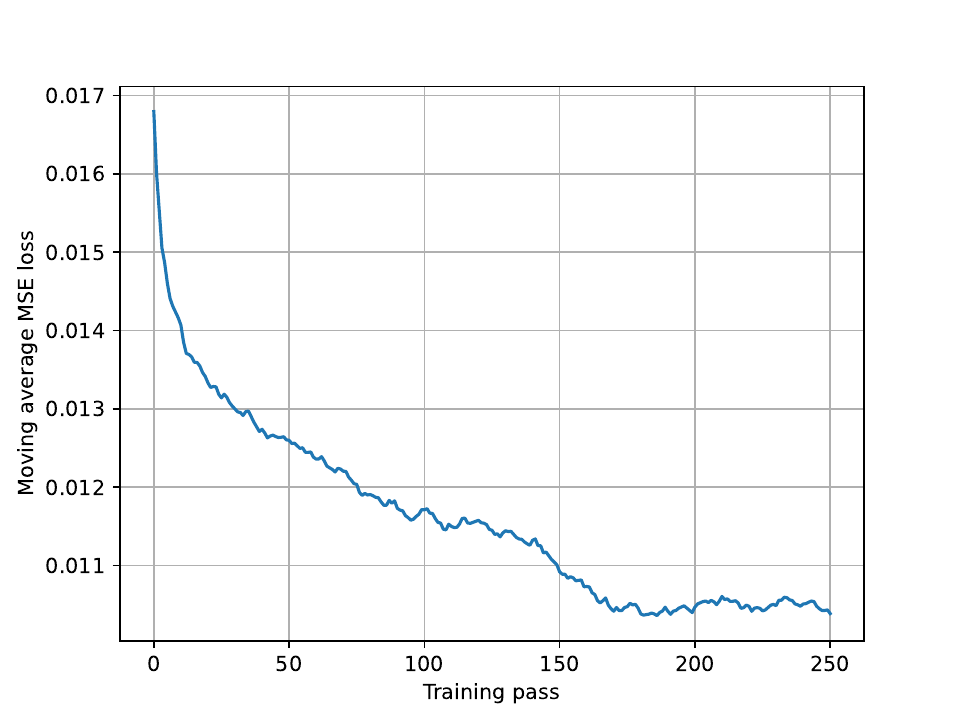}
        \caption{Single point functions moving average encoder network training loss (MNIST)}
        \label{fig:1_mnist}
    \end{subfigure}
    \hfill
    \begin{subfigure}[b]{0.4\textwidth}
        \centering
        \includegraphics[width=\textwidth]{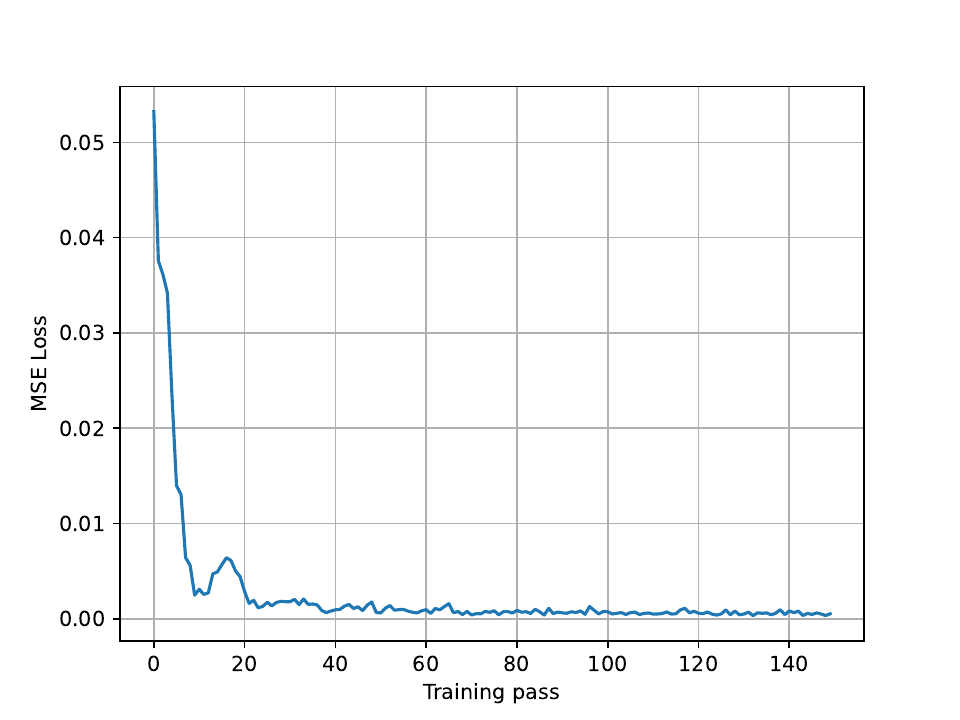}
        \caption{Single point functions decoder network training loss (MNIST)}
        \label{fig:2_mnist}
    \end{subfigure}
    
    \vspace{0.5cm}
    
    \begin{subfigure}[b]{0.4\textwidth}
        \centering
        \includegraphics[width=\textwidth]{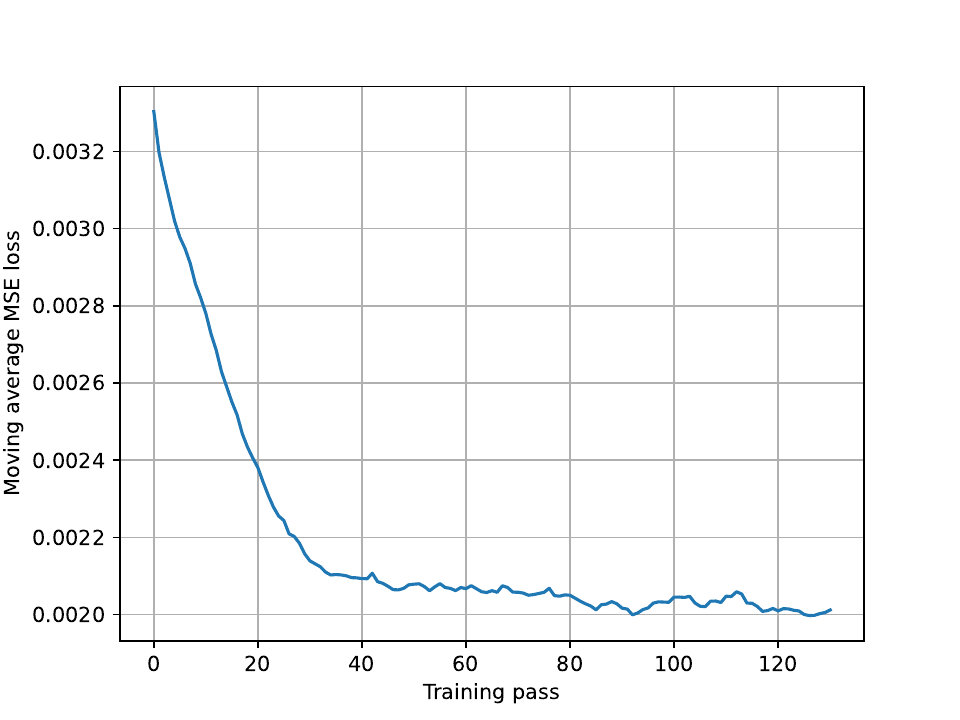}
        \caption{1- and 2-point function moving average encoder network training loss (MNIST)}
        \label{fig:3_mnist}
    \end{subfigure}
    \hfill
    \begin{subfigure}[b]{0.4\textwidth}
        \centering
        \includegraphics[width=\textwidth]{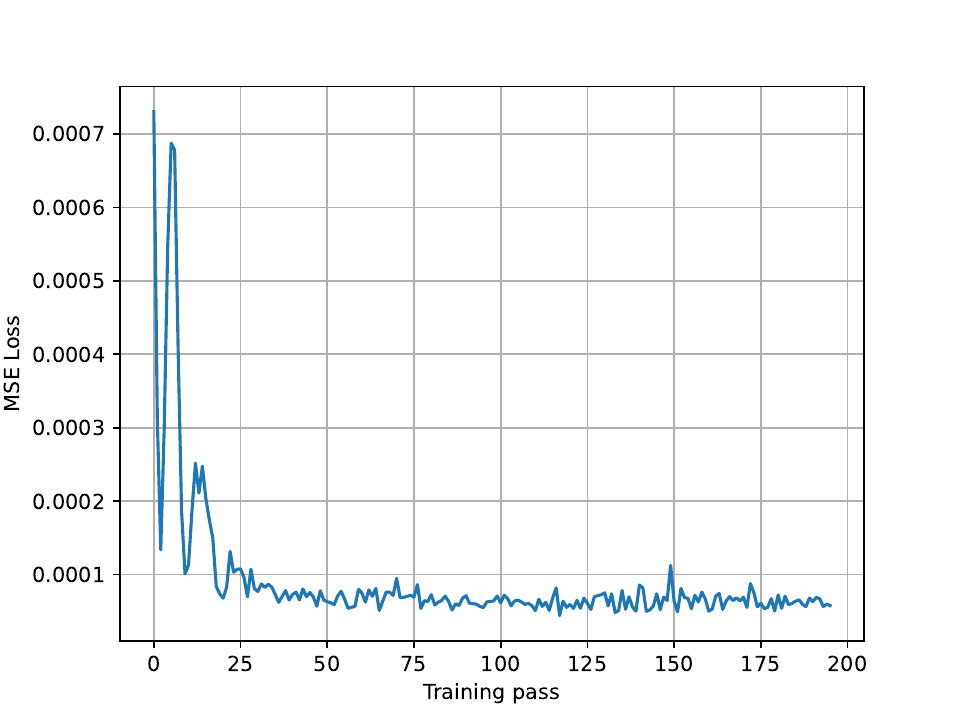}
        \caption{1- and 2-point function decoder network training loss (MNIST)}
        \label{fig:4_mnist}
    \end{subfigure}
    
    \vspace{0.5cm}
    
    \begin{subfigure}[b]{0.4\textwidth}
        \centering
        \includegraphics[width=\textwidth]{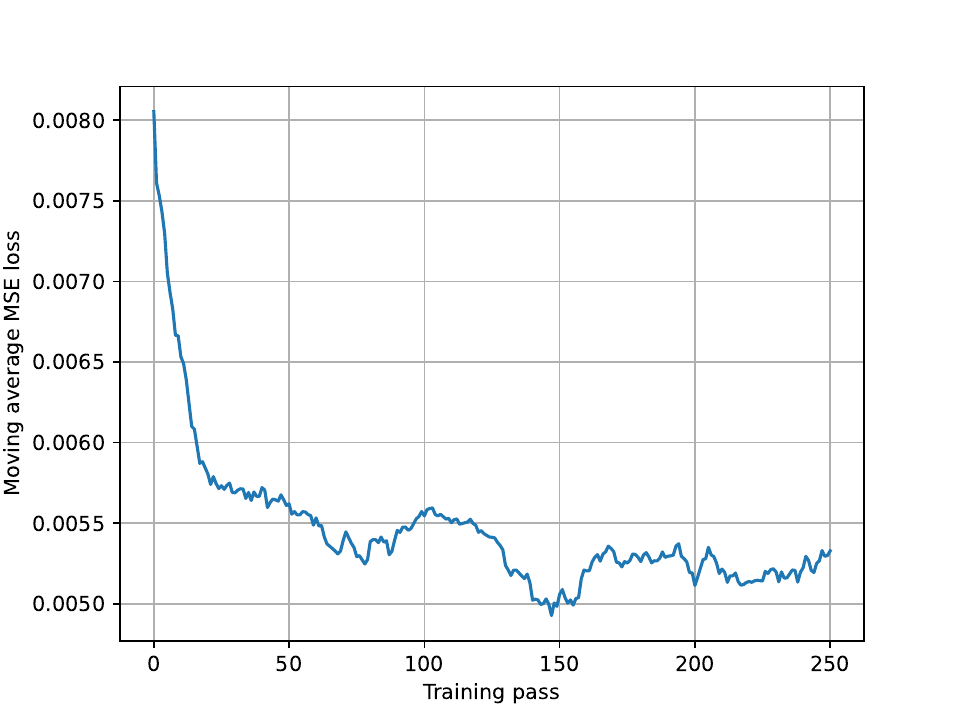}
        \caption{1-, 2-, and 3-point function encoder training loss (MNIST)}
        \label{fig:5_mnist}
    \end{subfigure}
    \hfill
    \begin{subfigure}[b]{0.4\textwidth}
        \centering
        \includegraphics[width=\textwidth]{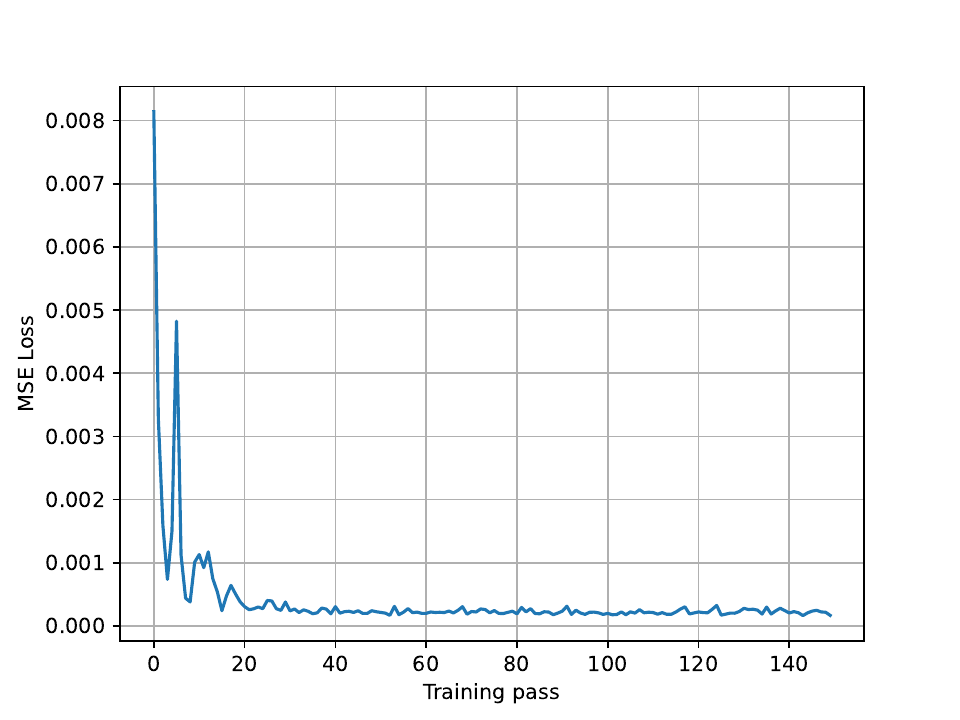}
        \caption{1-, 2-, and 3-point function decoder training loss (MNIST)}
        \label{fig:6_mnist}
    \end{subfigure}
    
    \caption{Training losses for MNIST dataset}
    \label{fig:mnist_grid}
\end{figure}

\begin{figure}[ht!]
    \centering
    
    \begin{subfigure}[b]{0.4\textwidth}
        \centering
        \includegraphics[width=\textwidth]{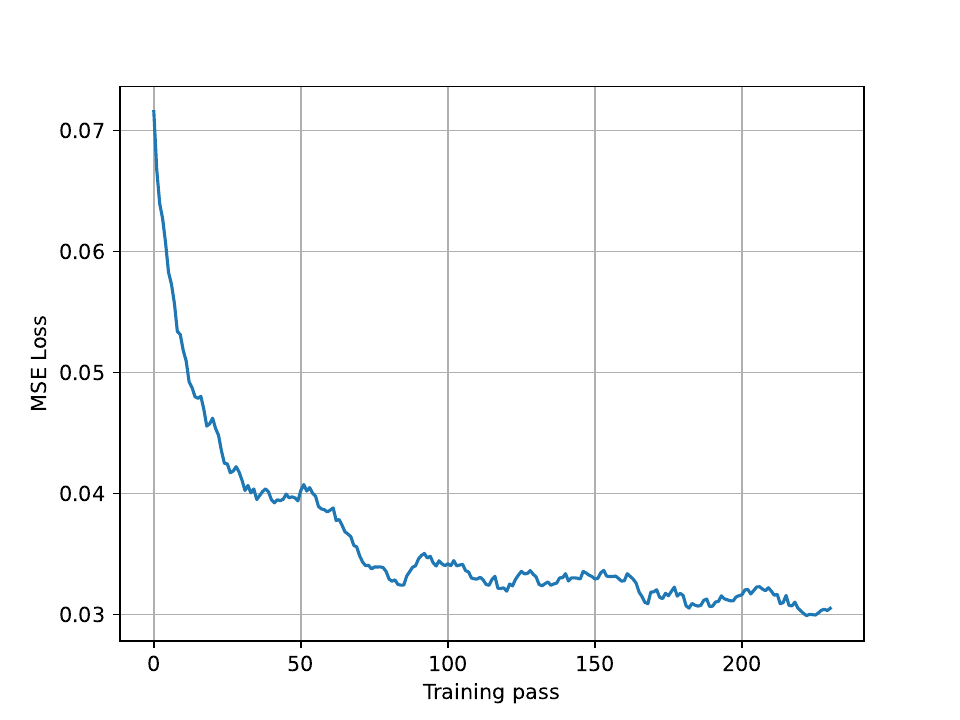}
        \caption{Single point functions moving average encoder network training loss (FashionMNIST)}
        \label{fig:1_fashion}
    \end{subfigure}
    \hfill
    \begin{subfigure}[b]{0.4\textwidth}
        \centering
        \includegraphics[width=\textwidth]{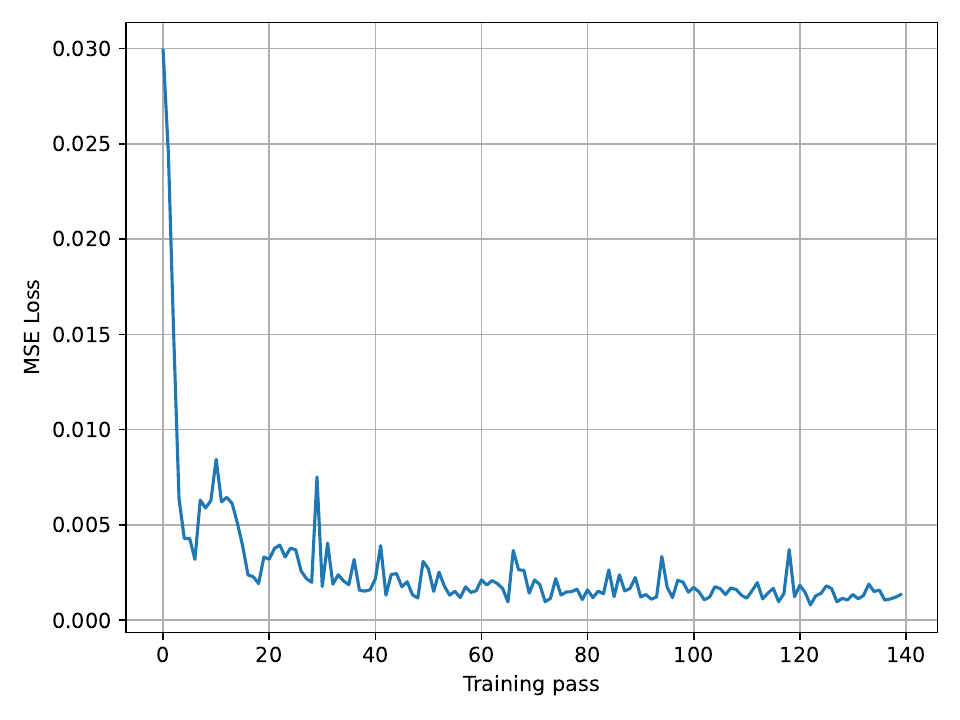}
        \caption{Single point functions decoder network training loss (FashionMNIST)}
        \label{fig:2_fashion}
    \end{subfigure}
    
    \vspace{0.5cm}
    
    \begin{subfigure}[b]{0.4\textwidth}
        \centering
        \includegraphics[width=\textwidth]{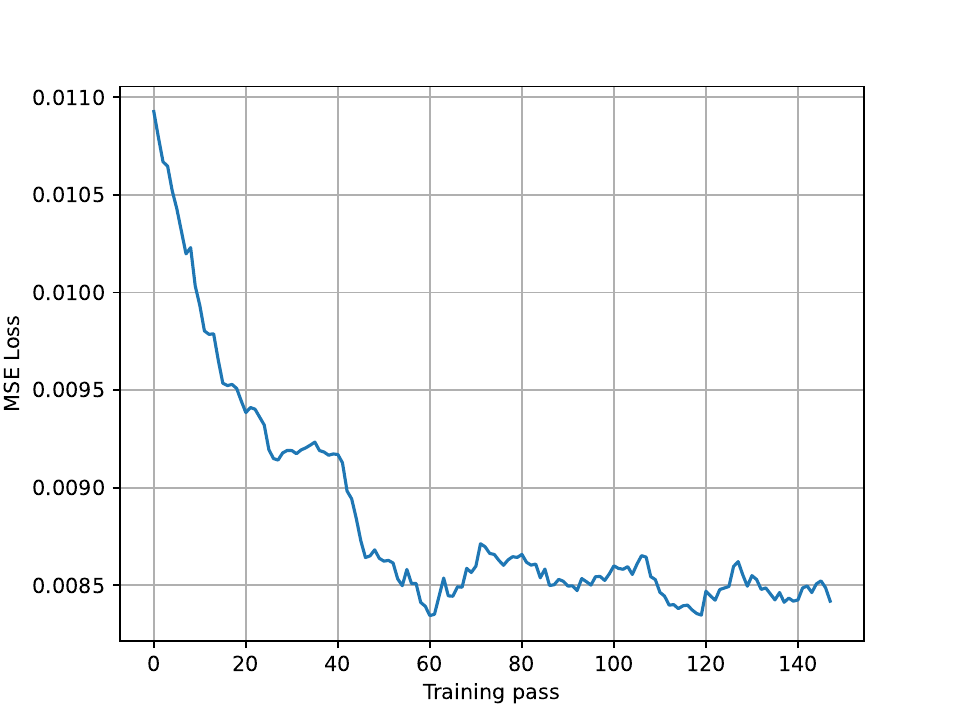}
        \caption{1- and 2-point function moving average encoder network training loss (FashionMNIST)}
        \label{fig:3_fashion}
    \end{subfigure}
    \hfill
    \begin{subfigure}[b]{0.4\textwidth}
        \centering
        \includegraphics[width=\textwidth]{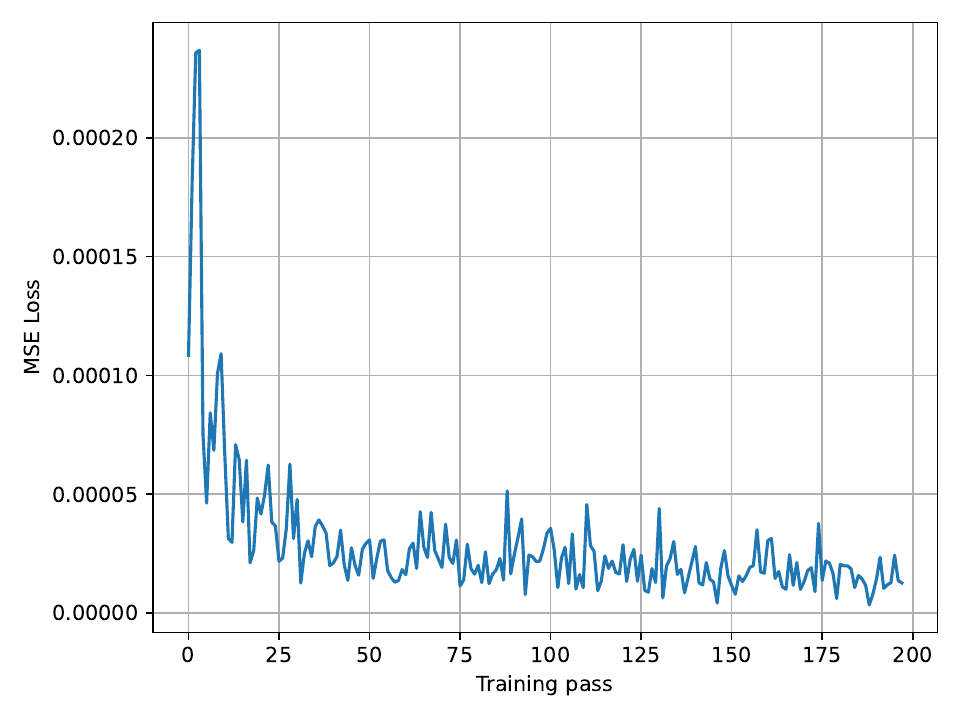}
        \caption{1- and 2-point function decoder network training loss (FashionMNIST)}
        \label{fig:4_fashion}
    \end{subfigure}
    
    \vspace{0.5cm}
    
    \begin{subfigure}[b]{0.4\textwidth}
        \centering
        \includegraphics[width=\textwidth]{figs/since_iaifi/final/FashionMNIST/mse_loss_encoder_1-all_2-0.8-3_0.008_avg100.pdf}
        \caption{1-, 2-, and 3-point function moving average encoder training loss (FashionMNIST)}
        \label{fig:5_fashion}
    \end{subfigure}
    \hfill
    \begin{subfigure}[b]{0.4\textwidth}
        \centering
        \includegraphics[width=\textwidth]{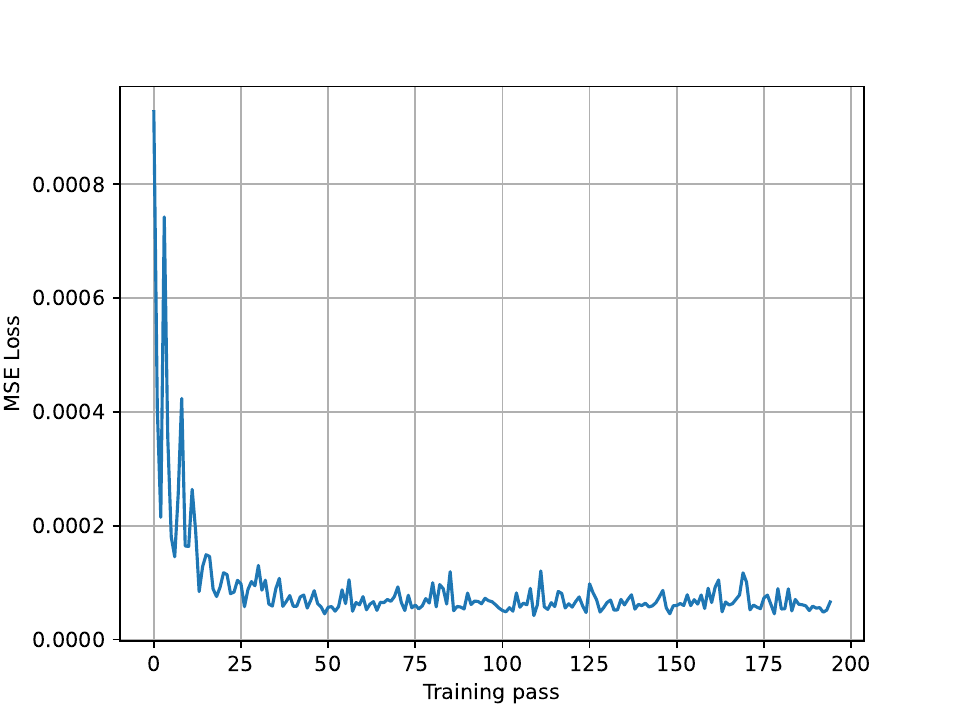}
        \caption{1-, 2-, and 3-point function decoder training loss (FashionMNIST)}
        \label{fig:6_fashion}
    \end{subfigure}
    
    \caption{Training losses for FashionMNIST dataset}
    \label{fig:fashion_grid}
\end{figure}

\section{Cumulants in Quantum Field Theory and Statistical Mechanics}%
\label{sec:_cumulants_in_quantum_field_theory}

In Section \ref{sec:statistical_prerequisites_and_cumulant_representations} we reviewed the statistics of distributions and the link between moments and cumulants; the following section highlights the links between quantum field theory and probability theory.

\subsection{Statistical Mechanics}%
\label{ssub:statistical_mechanics}

Consider the canonical ensemble encompasing all states of a thermodynamic system which is in thermal equilibrium with a fixed-temperature heat bath. The canonical ensemble has an associated probability distribution of states governed by the thermodynmaic variable of temperature $T$. For convenience, we introduce the `thermodynamic beta' $\beta = \frac{1}{k_B T}$, where $k_B$ is Boltzmann's constant. 

The probability $p_i := p(E = E_i)$ of being in an energy eigenstate $E_i$ is given by the Boltmann distribution
\begin{equation}
	p(E=E_i) = \frac{e^{-\beta E_i}}{Z[\beta]},
\end{equation}
where $Z[\beta]$ is the partition function. Notice that $Z[\beta]$ imposes that $\sum_i p_i = 1$.

The energy distribution has an associated moment-generating functional $M_E(t)$,
\begin{equation}
	M_E(t) = \sum_{i} p_i e^{tE_i} = \frac{1}{Z[\beta]} Z[\beta-t],
\end{equation}
which yields corresponding moments
\begin{equation}
	\frac{1}{Z[\beta]} \frac{\partial^k}{\partial t^k} Z[\beta-t] \bigg \rvert_{t=0} = (-1)^k \frac{Z^{(k)}[\beta]}{Z[\beta]},
\end{equation}
where the compact notation $Z^{(k)}[\beta] \equiv \frac{\partial^k}{\partial t^k} Z[\beta]$ has been introduced.

The associated cumulant generating functional is then
\begin{equation}
	\ln Z[\beta-t] - \ln Z[\beta],
\end{equation}
which proves interesting: the cumulants are independent of normalization, meaning one may instead take $\beta$ derivatives of the partition function and achieve the same result. Defining a new cumulant generating functional $\tilde K(\beta)$,
\begin{equation}
	\tilde K[\beta] := -\ln Z[\beta] = -\beta^{-1} F := - \tilde F
\end{equation}
meaning the cumulant generating functional is a multiple of the (Helmholtz) free energy\footnote{Notice our definition of the Helmholtz energy differs by a factor of -1 to the usual definition.} $F = \ln Z[\beta]$; the \textit{dimensionless free energy} $ \tilde F = \beta F$ serves as a cumulant generating functional. The corresponding $k$-th cumulant is thus
\begin{equation}
	\tilde \kappa_k = (-1)^{k-1} \tilde K^{(k)}[\beta].
\end{equation}

Further thermodynamic quantities such as the entropy and specific heat capacity can further be expressed in terms of the cumulants $\tilde \kappa_k$.

\subsection{Quantum Field Theory}%
\label{ssub:qft}

Consider a real scalar field $\phi$ in $d$ spacetime dimensions, and let $\phi$ be associated to a theory with Lagrangian density $\mathcal L$. The system has an associated \textit{partition functional}\footnote{The square brackets on the measure indicate a functional integral, and we choose to work in units where the reduced Planck constant $\hbar = 1$.}
\begin{equation}
	Z[J] = N \int [d\phi] \; \exp \bigg(i \int d^dx \; [\mathcal L + J(x) \phi(x)] \bigg),
\end{equation}
where $J(x)$ is the \textit{source} and $N$ a normalizing constant fixed to ensure the vacuum expectation value is unity. Taking functional derivatives\footnote{For those unfamiliar with the path integral formulation of QFT, we recommend the classic reference \cite{peskin2018introduction}, or for a gentler introduction \cite{cardy2010introduction}.} with respect to the source generates the $k$-point correlation functions,
\begin{equation}
	\braket{x_1, \ldots, x_k} = i^{-k} \frac{\delta^k Z[J]}{\delta J(x_1) \ldots \delta J(x_k)} \bigg \rvert_{J = 0}.
\end{equation}

Another quantity of interest in quantum field theory is the \textit{connected correlation function} $\braket{x_1, \ldots, x_k}_\text{Con.}$,
\begin{equation}
	\braket{x_1, \ldots, x_k}_\text{Con.} = i^{1-k} \frac{\delta^k W[J]}{\delta J(x_1) \ldots \delta J(x_k)} \bigg \rvert_{J = 0},
\end{equation}
where $W[J] := -i \ln Z[J]$ is the \textit{connected diagram generating functionl}.

As in \cite{Floerchinger:2023ekw}, one may generalize the discussion and instead change the random variables $\phi$ to depend on microscopic degrees of freedom $\gamma$ such that $\phi \to \phi[\gamma]$. 
\begin{figure}[htpb]
    \centering
    \begin{tabular}{|c|c|c|}
        \hline
        Quantity &  QFT & Statistical Mechanics\\
        \hline
        Source parameter & $J$ & $\beta = \frac{1}{k_B T}$\\
        Partition functional & $Z[J] \sim \int d[\phi] \, \exp (i \int d^dx \, [\mathcal L + J(x) \phi(x)])$ & $Z[\beta] \sim \sum_i e^{-\beta E_i}$\\
        Cumulant generating functional & $W[J] = -i \ln Z[J]$ & $\tilde K[\beta] = - \ln Z[\beta] $\\
    $k$-th Moment & $i^{-k} \frac{\delta Z[J]}{\delta J(x_1) \ldots \delta J(x_k)} \big \rvert_{J=0}$ & $(-1)^{k-1} \tilde K^{(k)} [\beta]$\\
        \hline
    \end{tabular}
    \caption{Dictionary between corresponding properties in QFT and Statistical Physics.}%
    \label{fig:qft_stat_phys_table}
\end{figure}

\section{Cumulant space equivalence classes}%
\label{sub:cumulant_space_equivalence_classes}

One useful notion in the study of cumulant-space autoencoders is that of equivalence; two useful notions are those of cumulant-space equivalency and sample-space equivalency. 

As an example of sample-space equivalency, one may construct a class of all cumulant vectors which contain sufficient information to produce identical samples down to some precision $\epsilon$, i.e.,
\begin{equation}
	\label{eqn:suff_class}
  K_\text{suff} = \bigg\{ K^{\tmi, \beta} \; \biggm|\;  \sum_j \sum_i |B_{j,i}^{\beta} - D_\phi (K^{\tmi, \beta})_i | ^2 < \epsilon  \; \forall \; \beta \bigg \},
\end{equation}
for optimal parameters $\phi_k$. These we define to be a \textit{sufficiency class}.

Alternatively, an equivalence class can also be defined such that the cumulants of the decoded samples are identical to some precision, but not necessarily sufficient. This may have the physical interpretation of an insufficient set of cumulants which flow to the same fixed point, down to some precision $\epsilon$, however not the original theory. 

Let the cumulants of a decoded, potentailly sampled batch be denoted by the vector $\Gamma^{\tmi, \beta}$, where, for some fixed but arbitrary ordering of $\tilde{\mathcal I}^{(k)}$,
\begin{equation}
	\label{eqn:def_of_gamma}
	\Gamma^{\tmi, \beta} := \bigoplus_{k=k_i}^{k_f} \big( \kappa_\beta(\zeta_{i_1} \ldots \zeta_{i_k}) \; \bigm| \; i_1, \ldots, i_k \in I, \, I \in \tilde{\mathcal I}^{(k)} \big),
\end{equation}
$\kappa_\beta$ denotes the numerical cumulants evaluated over the batch $\beta$, and $\zeta_i = (D_\phi E_\theta(B^{\beta}))_i$.

A cumulant space equivalency class is thus given by
\begin{equation}
	\label{eqn:equiv_classes}
K_\text{eq} = \big\{K^{\tmi, \beta} \; \bigm|\;  |\Gamma^{\tmi,  \beta} - K^{\tmi, \beta}|^2 < \epsilon  \; \forall \; \beta \big \},
\end{equation}
where in general $K_\text{eq} \supseteq K_\text{suff}$.

\section{Classifiers}%
\label{sec:classifiers}

Consider a classifier network denoted by $C$, 
\begin{align}
	C: \chi &\to \mathbb R^c\\
	z &\mapsto \delta
\end{align}
where $c$ is the number of classes.

We propose two possible values of $c$ offering differing physical interpretations. Arguably the simplest approach, corresponding to $c=2$, is to train the classifier on a superset of the training data of the autoencoder, with the labels $0$ or $1$ corresponding to whether the data is contained within $T$ or not. In prior analysis, namely that contained in Section \ref{sec:reconstruction}, we considered the case where each data sample is the IID realization of some distribution over the samples; in the case of the analysis of the MNIST dataset, this means we only took one digit type (e.g. sevens). The entire MNIST dataset $P$ (constituting all digits)\footnote{$P$ is the \textit{parent} dataset $P \supset T$.} can be de-composed into a pair of \textit{disjoint} sets,
\begin{equation}
	P = T \cup T'
\end{equation}
where $T' = \{ z_j\}$ contains data which is dimensionally identical to $ y_j \in T$, but not in the autoencoder training set. More concretely, in our analysis $T$ contains all realizations of the digit `7' from the MNIST set, and $T'$ contains all other digits.

Alternatively, for datasets such as MNIST and FashionMNIST which contain different classes of sample, one can consider the case where $c = \text{number of classes}$. Rather than determining whether the sample is in the training dataset or not, one may consider the case where $P$ is instead decomposed into classes $(c_1, \ldots, c_l)$,
\begin{equation}
	P = T \cup T_1 \cup \ldots \cup T_l,
\end{equation}
where once more $T_c = \{  z_j \}$ contains data this is dimensionally identical to that in $T$. This choice corresponds to $c = l + 1$ classes, and is a more thorough measure of how well the autoencoder has learnt the features of the dataset as the classifier has more choices of distribution to choose from.

The classifier is trained on $P$, with a loss function such as \textit{cross-entropy loss} 
\begin{align}
	\mathcal L_\text{classification}(x,y)  = (l_1, \ldots, l_c)\\
  l_n = - \sum_\rho^c \log \frac{e^{x_{n,\rho}}}{\sum_i^c e^{x_{n, i}}} y_{n,\rho}
\end{align}
such that
\begin{equation}
	\text{max}(\delta) = \begin{cases}
		1 \quad  z \in T\\
		0 \quad \text{otherwise}.
	\end{cases}
\end{equation}
Training this classifier with $P$ as the entire dataset allows one to benchmark how well the output of the cumulant network encodes the information in the dataset. The higher the validation accuracy, the better the networks trained on $n$-point functions encode the samples. We emphasise that the specific architecture of the classifier is unimportant provided it yields a sufficiently good validation accuracy when acting on unseen test data; the network will be a poor measure of similarity if it has begun to over-fit.

\newpage

\bibliography{cumulant_encoder}
\bibliographystyle{JHEP}

\end{document}

%% file: figs/tikz/map_lattice.tex
\begin{tikzpicture}[scale=1.2]

\foreach \i in {0,...,8}
{
  \pgfmathtruncatemacro{\x}{mod(\i, 3)}
  \pgfmathtruncatemacro{\y}{int(\i / 3)}
  \draw (\x,\y) circle (2pt) node[below] {$x_{\i}$};
}

\foreach \x in {0,1,2}
  \foreach \y in {0,1}
    \draw (\x,\y) -- (\x,\y+1);

\foreach \x in {0,1}
  \foreach \y in {0,1,2}
    \draw (\x,\y) -- (\x+1,\y);

\draw[->, thick] (2.5,1) -- node[above] {$\Phi$} (3.5,1);

\foreach \i in {0,...,8}
{
  \pgfmathtruncatemacro{\x}{mod(\i, 3) + 4}
  \pgfmathtruncatemacro{\y}{int(\i / 3)}
  \filldraw[black] (\x,\y) circle (2pt) node[below] {$y_\i$};
}

\foreach \x in {4,5,6}
  \foreach \y in {0,1}
  \filldraw[black] (\x,\y) -- (\x,\y+1);

\foreach \x in {4,5}
  \foreach \y in {0,1,2}
  \filldraw[black] (\x,\y) -- (\x+1,\y);

\node[anchor=east] at (-0.5,1) {$L$};
\node[anchor=west] at (6.5,1) {$y_i = \Phi(x_i) \in S$};

\end{tikzpicture}

%% file: figs/tikz/reconstruction.tex
    \begin{tikzpicture}[
        node/.style={draw, circle, minimum size=0.6cm},
        arrow/.style={-, shorten >=1pt, >=stealth, semithick}
    ]
        \foreach \x [count=\xi] in {1,...,4} {
            \node[node] (input\x) at (0, -\xi) {$x_{\x}$};
        }
				\node[above=0.1cm of input1] {Input};
        
        \foreach \x [count=\xi] in {1,...,2} {
            \node[node] (hidden\x) at (2.5, -\xi-1.0) {$h_{\x}$};
        }
				\node[above=0.1cm of hidden1] {Latent space};
        
        \foreach \x [count=\xi] in {1,...,4} {
            \node[node] (output\x) at (5, -\xi) {$y_{\x}$};
        }
				\node[above=0.1cm of output1] {Output};
        
        \foreach \x in {1,...,4} {
            \draw[arrow] (input\x.east) -- (hidden1.west);
            \draw[arrow] (input\x.east) -- (hidden2.west);
        }
        
        \foreach \x in {1,...,2} {
            \foreach \y in {1,...,4} {
                \draw[arrow] (hidden\x.east) -- (output\y.west);
            }
        }
    \end{tikzpicture}